\documentclass{article}
\usepackage{amsmath,amssymb}
\usepackage{epsfig,multirow}
\usepackage{array}
\usepackage[a4paper,top=3cm,bottom=2cm,left=3cm,right=3cm,marginparwidth=1.75cm]{geometry}
\title{
Anisotropies of  ultrahigh-energy cosmic rays \\in a scenario with nearby sources}

\author{Silvia Mollerach and Esteban Roulet\\
Centro At\'omico Bariloche, Comisi\'on Nacional de Energ\'\i a At\'omica\\
Consejo Nacional de Investigaciones Cient\'\i ficas y T\'ecnicas (CONICET)\\
Av. Bustillo 9500, R8402AGP, Bariloche, Argentina}
\date{}

\begin{document}
\maketitle
\begin{abstract}
    The images of ultrahigh-energy cosmic ray sources get distorted, in an energy dependent way, by the effects of Galactic and extragalactic magnetic fields. These deflections can also affect the observed cosmic ray spectrum, specially when the sources are transient. We study scenarios in which one or a few nearby extragalactic sources, such as Cen~A or M81/M82,  provide the dominant contribution to the cosmic ray flux above the ankle of the spectrum. We discuss the effects of the angular dispersion induced by the turbulent extragalactic magnetic fields, and the coherent deflections caused by the regular Galactic magnetic field, with the associated multiple imaging of the sources. We consider the possible contribution from those sources to the dipolar distribution discovered by the Pierre Auger Observatory above 8~EeV, as well as to the hot spots  hinted in the observations by the Pierre Auger and Telescope Array observatories at higher energies, taking into account the mixed nature of the cosmic ray composition.
\end{abstract}
\section{Introduction}
The sources of the cosmic rays (CRs) are still unknown, and although it is believed that below $10^{17}$\,eV they are mostly Galactic, probably associated with supernova remnants or with pulsars, at the highest energies the CRs are most likely of extragalactic origin.  The indications for this come mostly from the limitations of Galactic sources  to reach energies well in excess of $10^{18}$\,eV, the lack of an anisotropy along the Galactic disk (which  would be expected at the highest energies if CR sources were Galactic) \cite{ls12}, and the observed dipolar anisotropy above 8\,EeV, pointing to a direction which is more than 100$^\circ$ away from the Galactic center direction \cite{science}.  Also hints of more localized  overdensities, on angular scales of about 20$^\circ$, have been reported by the Pierre Auger \cite{augerhs} and Telescope Array \cite{tahs} observatories at energies above about 40\,EeV, and they are not associated with directions along the Galactic plane. 

Different scenarios for extragalactic sources have been considered, being the traditional one based on a cosmological distribution of powerful sources, such as active galactic nuclei, gamma ray bursts, tidal disruption events, galaxy mergers, etc. The evolution with redshift of the luminosity density of these sources is a relevant ingredient, affecting both the attenuation of the CR fluxes at the highest energies as well as the production of secondaries by the interactions with the radiation backgrounds (including the production of nuclear fragments in the photodisintegration of heavier primary nuclei or the production of photons and neutrinos in photo-pion interactions). In this kind of scenarios, the dipolar anisotropies observed by the Pierre Auger Observatory are naturally related to the non-uniform distribution of the galaxies in the local neighborhood within few hundred Mpc, with the CR arrival directions being also affected by the deflections induced by the Galactic magnetic field \cite{apj18}. 
The relevance of the presence of nearby sources to explain the CR spectrum at the highest energies, and eventually also to account for the observed anisotropies on intermediate angular scales, has been studied, e.g.,  in refs.~\cite{bl99,ta11,ma18,la20}.

Scenarios in which some of the nearby sources provide a sizeable fraction of the overall fluxes, giving rise to the more localized overdensities observed by the Pierre Auger Observatory above 40\,EeV, have been compared with observations,  with the conclusion that about 10\% of the events at these energies could be localized in regions of angular radius of about 20$^\circ$ around some nearby sources. These could be for instance specific starburst galaxies or active galaxies, such as Centaurus~A. The remaining $\sim 90$\%  of the flux reaching the Earth, being more isotropically distributed, could be attributed in this case to the sources which lie farther away \cite{augerhs}.

A different possibility is that just one or two powerful nearby sources provide most of the flux above the energy  of the ankle of the spectrum (which is the hardening observed at 5\,EeV). In this kind of scenario \cite{mo19}, the rather isotropic CR distribution observed up to high energies  is attributed to the almost diffusive propagation of the heavier component of the CRs emitted by the same nearby sources. The diffusive propagation would result from the presence of turbulent magnetic fields in the local supercluster region \cite{va11,fe12}, which could significantly deflect the CR trajectories, specially for those  heavier components.  Moreover, relatively low rigidity cutoffs at the sources, of order 10\,EV,  can allow to avoid the appearance of overdensities on small angular scales around the source directions, as would be expected if the light components were to extend up to the highest energies.  An additional ingredient  is related to the possible non-steadiness  of the sources, as could happen in the case in which the emission is associated to a strong burst or to a recent period of  activity, resulting for instance from the effects of galaxy-galaxy interactions. In this case, one also expects a strong signal to be imprinted in the observed CR spectrum, since low-energy CRs may take longer than the age of the source to reach the Earth, and  hence a strong suppression of the observed source spectrum would be expected to take place at low rigidities. This feature can be useful to account for the inferred shapes of the spectrum of the different CR mass components above the ankle, which indeed look very hard at low rigidities \cite{combinedfit}. 

In this work we want to study in detail how a nearby source in such a scenario would appear to an observer at the Earth. We want in particular to discuss the main properties of the source and of the extragalactic magnetic field that would be required in order to give rise to features in the spectrum and arrival directions distribution compatible  with those actually observed. To be specific, we will consider as illustrative sources the cases of Centaurus~A and that of the M81/M82 galaxy pair. These sources are quite nearby, being at approximate distances of 3.8 and 3.6~Mpc respectively, and they provide attractive candidates of ultrahigh-energy CR accelerators. The giant elliptical radiogalaxy Cen~A \cite{is98} is actually the result of a merger of two galaxies, which happened some $10^8$--$10^9$\,yrs ago. It  harbors a supermassive black hole \cite{cenaBH} with a mass of $\sim 5\times 10^7\,M_\odot$ and has two powerful outer jets with radiolobes as well as inner jets. The inner lobe may be the result of the most recent, and still active, outburst,  while the outer lobes may be the result of previous activity that took place several hundred million years ago. The galaxy M81 is a giant spiral, with a central black hole mass of $7\times 10^7\,M_\odot$ \cite{m81BH}, being a Seyfert\,I, although with weak activity at present. It interacted with the M82 galaxy some few hundred million years ago, what probably induced the strong starburst activity of this last, and the two galaxies are deemed to eventually  merge into a single one in future interactions (the smaller galaxy NGC~3077 contributes to actually form a trio).  M82 has some AGN activity, with a central black hole of $3\times 10^7\,M_\odot$. Note that while Cen~A is almost at the center of the hot-spot hinted by Auger observations above 40\,EeV \cite{augerhs}, the M81/M82 complex is not far from the location of the Telescope Array reported hot-spot above about 57\,EeV \cite{tahs}.

\section{The nearby source scenario}

In this section we describe the main characteristics relevant for the nearby source scenario of ref.~\cite{mo19} that will be considered in this work. When CRs of atomic number $Z$ emitted from a steady source at a distance $r_{\rm s}$ propagate  through a homogeneous turbulent random extragalactic field of root mean square strength $B_{\rm rms}$ and coherence length $l_{\rm c}$, the distribution of arrival directions at the Earth (ignoring energy losses and the deflections caused by the Galactic magnetic fields) is well described by a Fisher distribution of the form
\begin{equation}
    \frac{1}{N}\frac{{\rm d}N}{{\rm d}\cos\theta}=\frac{i}{2} +(1-i)\frac{\kappa\exp(\kappa \cos \theta)}{2 \sinh \kappa} ,
    \label{eqfisher}
\end{equation}
where $\theta$ is measured with respect to the source direction and the concentration parameter $\kappa$ is approximately given by the expression
\begin{equation}
    \kappa^{\rm (steady)}\simeq \frac{1}{R_{\rm s}}\left[2+\exp(-2R_{\rm s}/3-R_{\rm s}^2/2)\right],
\end{equation}
with $R_{\rm s}\equiv r_{\rm s}/l_D$ being the source distance in units of the diffusion length $l_D$ \cite{hmr16}. This last characterizes the distance after which the average deflections in the random field become of order 1~radian, and is hence energy dependent. An approximate expression for the diffusion length, in the case that we will consider of a Kolmogorov spectrum of turbulence, is \cite{hmr14}
\begin{equation}
    l_D\simeq l_{\rm c}\left[4\left(\frac{E}{E_{\rm c}}\right)^2+0.9\frac{E}{E_{\rm c}}+0.23\left(\frac{E}{E_{\rm c}}\right)^{1/3}\right],
\end{equation}
with the critical energy (for which the effective Larmor radius equals the coherence length) being $E_{\rm c}=eZB_{\rm rms}l_{\rm c}\simeq 0.9\,{\rm EeV}(B_{\rm rms}/{\rm nG})(l_{\rm c}/{\rm Mpc})$. The  isotropic term in eq.~(\ref{eqfisher}) involving the parameter $i$ arises from particles that made several turns before reaching us, and one has that \cite{hmr16}
\begin{equation}
    i\simeq 1-\frac{\langle\cos\theta\rangle}{{\rm coth}\kappa-1/\kappa},
\end{equation}
where the average cosine of the deflections is approximately
\begin{equation}
    \langle\cos\theta\rangle^{\rm (steady)}\simeq \frac{1}{3R_{\rm s}}\left[1-\exp(-3R_{\rm s}-3.5R_{\rm s}^2)\right]\equiv C(R_{\rm s}).
    \label{eqcos}
\end{equation}
At high enough rigidities, when $l_D>r_{\rm s}$ (or $R_{\rm s}<1$), the propagation will become almost rectilinear, and one may introduce the effective average deflection $\bar\theta$ through $\langle\cos\theta\rangle\simeq 1-\bar\theta^2/2$.
This deflection satisfies
\begin{equation}
    \bar\theta\simeq \sqrt{\frac{r_{\rm s}}{6l_{\rm c}}}\frac{E_{\rm c}}{E}\simeq 20^\circ\sqrt{\frac{r_{\rm s}}{\rm 4\,Mpc}\frac{l_{\rm c}}{\rm 30\,kpc}}\frac{B_{\rm rms}}{\rm 100\,nG}\frac{\rm 40\,EeV}{E/Z},
    \label{eqtheta}
\end{equation}
where we have normalized the parameters to typical values to be considered later on. 
Note that if the magnetic field strength in the local neighborhood of few Mpc is significant ($B_{\rm rms}\sim 10$ to 100\,nG), a 10\,EeV CR proton will typically be deflected by more than 10$^\circ$ along its trajectory, and heavier CRs would be deflected by an amount larger by a factor $Z$. Given that the CRs will reach the observer through different magnetic field domains, the random deflections should lead to a blurring of the images of the sources.

At much lower rigidities, i.e. for $R_{\rm s}\gg 1$, the CRs will  suffer strong deflections before reaching the Galaxy, having a diffusive behavior that will lead to small departures from an isotropic distribution, with the dipolar component having an amplitude $\Delta\simeq 3\langle\cos\theta\rangle\simeq 1/R_{\rm s}$.\footnote{Let us note that the inclusion of energy losses  would enhance $\langle\cos\theta\rangle$ (and hence $\Delta$), with the effect becoming increasingly important for more distant sources \cite{hmr16}. However, for close-by sources the effects are small, specially if the emission only started at redshifts much less than unity.}

This diffusive propagation will also lead to an enhancement $\xi$ in the CR density $N$, with respect to the result that would be obtained in the case of rectilinear propagation, such that
\begin{equation}
    N(r)\equiv \xi\frac{Q}{4\pi c r^2},\ \ \ \ {\rm with}\ \ \ \ \xi=\frac{1}{\langle\cos\theta\rangle},
\end{equation}
so that the total flux through a sphere of radius $r$, $\Phi=4\pi r^2 c\langle\cos\theta\rangle N(r)$, remains equal to the source emissivity $Q$, as expected for a steady source \cite{mo19}. This local density enhancement is quite sizeable when the distribution of arrival directions becomes close to isotropic, since  the overdensity is
related to the dipolar amplitude through $\xi\simeq 30(0.1/\Delta$).

\begin{figure}[t]
    \centering
    \includegraphics[width=0.49\textwidth]{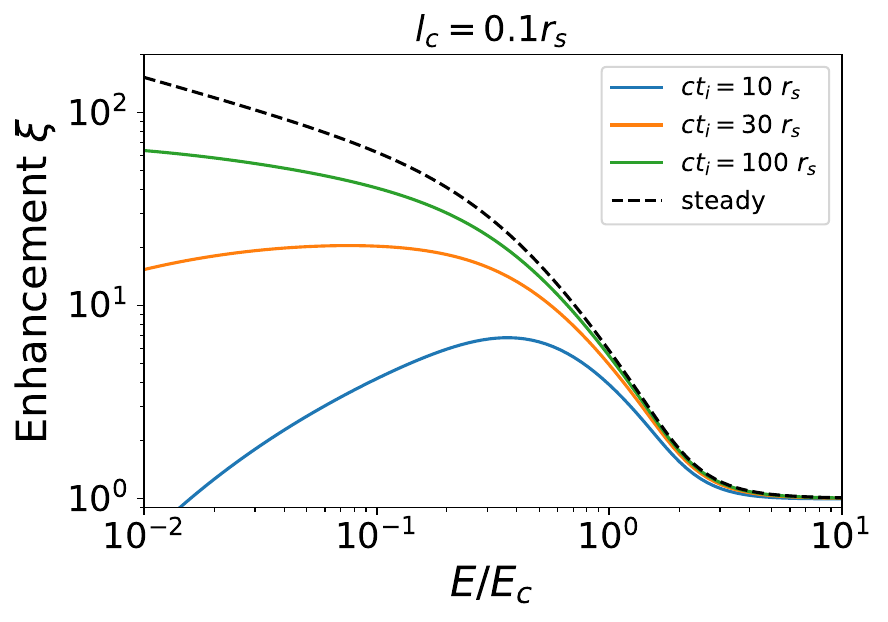}\includegraphics[width=0.49\textwidth]{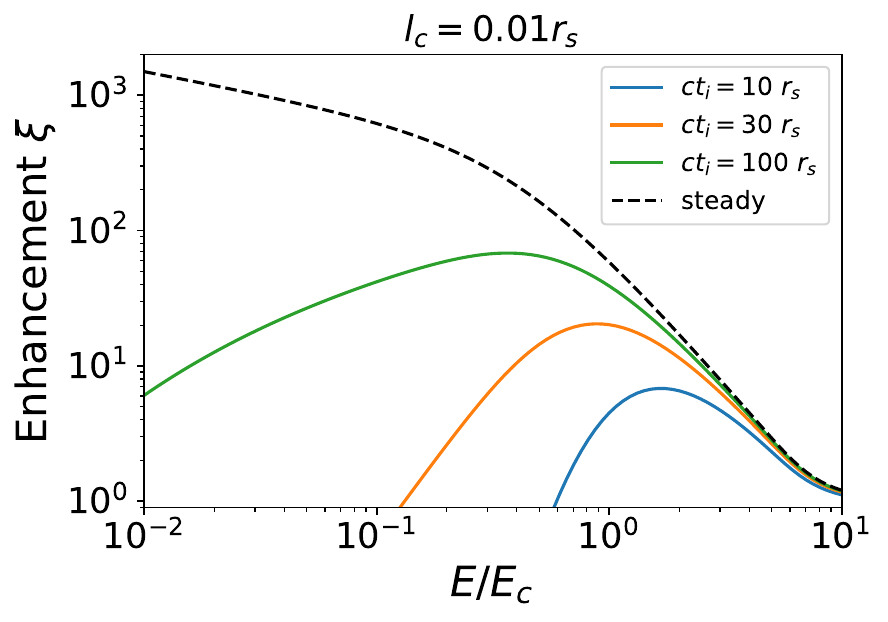}
    \caption{Spectral enhancement factor for transient sources emitting steadily since different initial times $t_i$ and for different values of $l_{\rm c}/r_{\rm s}$.}
    \label{fig1}
\end{figure}

If the source is instead transient, some additional characteristics may appear in the spectrum and anisotropies. In particular, if the CR emission started a time $t_i$ before present, the CR density will be strongly suppressed beyond a distance $r\simeq \sqrt{ct_i l_D}$ and, correspondingly, the observed spectrum from the source at distance $r_{\rm s}$ will be strongly suppressed below the energy for which $l_D\simeq r_{\rm s}^2/ct_i$.
Specifically, one finds that for a source starting to emit CRs at a time $t_i$ and emitting steadily afterwards, the enhancement factor of the observed spectrum takes the form \cite{mo19,ha21}
\begin{equation}
    \xi_i\simeq \frac{1}{C(r_{\rm s}/l_D)}\exp\left[-\left(\frac{r_{\rm s}^2}{0.6l_Dct_i}\right)^{0.8}
    \right],
\end{equation}
with $C(R)$ given in eq.~(\ref{eqcos}). The overall shape of the enhancement is displayed in Figure~\ref{fig1} as a function of $E/E_{\rm c}$ for different values of $ct_i/r_{\rm s}$ and for $l_{\rm c}=r_{\rm s}/10$ (left panel) and $l_{\rm c}=r_{\rm s}/100$ (right panel). Further corrections to these expressions appear when $ct_i\simeq r_{\rm s}$, but we do not consider those cases since in that situation the more isotropic low-energy contribution will not be present.

The main salient features of the results displayed in Figure\,\ref{fig1} are:
\begin{itemize}
    \item the enhancement starts to be noticeable when the propagation becomes non-rectilinear, i.e. for $l_D<r_{\rm s}$, which corresponds to $E< E_{\rm rect}=E_{\rm c}\sqrt{r_{\rm s}/l_{\rm c}}$
    
    \item for the steady case, the enhancement $\xi$ scales approximately as $E^{-2}$ for $E/E_{\rm c}>1$ and as $(E/E_{\rm c})^{-1/3}$ for $E/E_{\rm c}<0.1$, following the energy dependence of $R_{\rm s}=r_{\rm s}/l_D$, since in the diffusive regime $\xi\simeq 3R_{\rm s}$.
    
    \item for finite initial emission times, $\xi$ reaches a maximum value at an energy $E_{\rm max}\simeq E_{\rm c} r_{\rm s}/\sqrt{4ct_il_{\rm c}}$, which can be obtained in the approximation that the diffusion length at high energies is $l_D\simeq 4l_{\rm c}(E/E_{\rm c})^2$
    
    \item the suppression at energies below $E_{\rm max}$ is very pronounced if $E_{\rm max}>0.5E_{\rm c}$, 
    but it is less pronounced if the maximum happens at lower energies, already in the resonant diffusion regime in which the energy dependence of the diffusion length is milder.

\end{itemize}

Regarding the distribution of arrival directions, given that there are no CRs emitted before $t_i$, which would have contributed in a relatively more isotropic way, the anisotropies are expected  to be enhanced with respect to the results for a steady source, specially at lower energies (see \cite{mo19,ha21} for details). In particular, one finds that in the diffusive regime and for $ct_i\gg r_{\rm s}$
\begin{equation}
    \langle\cos\theta\rangle\simeq  C(R_{\rm s})\left[ 1+\left(1.6-\frac{3}{2d_i}\right)\left( \frac{R_{\rm s}^2}{0.7d_i}\right)^{0.8-0.5/d_i}\right],
    \label{eqcosi}
\end{equation}
with $d_i\equiv ct_i/l_D$. Regarding the concentration parameter, one has
\begin{equation}
    \kappa\simeq \kappa^{\rm (steady)}+\frac{0.44}{(d_i/R_{\rm s})^{0.8+0.4/d_i}-1}.
\end{equation}

A scenario with a bursting source, in which the emission takes place just in a narrow time interval, has some interesting peculiarities. In particular, since  regardless of their energy all the CRs travelled the same amount of time from the source to the observer, their dipolar amplitude in the diffusive regime turns out to be energy independent \cite{be90}. However, the observable spectrum of the source is largely affected, being peaked in a narrow energy range \cite{mi96}.  A very strong suppression is expected  at high energies, when the typical time for  propagation from the source becomes shorter than the time since the burst happened. The lack of emission of CRs at times posterior to the burst, which could have arrived through straighter trajectories, reduces the dipolar amplitudes with respect to the expectations from a source that would have emitted steadily since the time of the burst. Since we will be interested in considering scenarios in which an individual source can contribute both to an intermediate angular scale localized anisotropy at the highest rigidities as well as to a more isotropic background at lower rigidities, we will consider in the following the case of continuous emission since some given initial time rather than the bursting case.

The features described above were exploited in ref.~\cite{mo19} to account for the observed spectrum, composition and large-scale anisotropies in a scenario in which the CR fluxes above few EeV were dominated by just one nearby source, such as Centaurus~A (or few sources with similar characteristics). In particular, the diffusive enhancement effect could help to explain (eventually in combination with a source rigidity acceleration cutoff) the high-energy suppression of the flux from each mass component, while the suppression of the flux at low energies due to the finite source emission time could account for the inferred pronounced hardness of the component spectra, even if the source spectra are consistent with the expectations from diffusive shock acceleration (i.e., with an approximate $E^{-\gamma}$ shape, with $\gamma\simeq 2$ to 2.4).
The observed dipole amplitudes, of order 5\% at 10~EeV,  could also result naturally from the more isotropic component of the heavier nuclei, while the lack of very localized anisotropies at the highest energies  could be associated to the strong suppression of the lighter components in this regime.

\section{Effects of Galactic magnetic field deflections}

The effect on the CR arrival directions (AD) of the turbulent extragalactic magnetic field described above is to give rise to a smeared AD distribution outside the Galaxy, which  depends on the CR rigidity, the magnetic field parameters and also on the temporal history of the source emission. This  distribution will be subsequently modified by the Galactic magnetic field induced deflections, leading to a more complex observed distribution at the Earth \cite{ha10}.

In this work we want to study in detail the effects of the Galactic magnetic field deflections in a scenario of this kind, involving just a couple of nearby sources,  which are taken for definiteness to be Cen~A and M81/M82.   We want  to explore also the possibility to account for the hints of anisotropies on intermediate angular scales \cite{augerhs,tahs} as being produced, in the scenarios considered, by the high-energy tail of the lighter CR components from those sources.

The Galactic magnetic field has both  regular and turbulent components, which are localized along the disk but also extend beyond it \cite{ha15}. Their strength near our location is of a few $\mu$G, and while the regular field is coherent over scales of several kpc, the random one has a coherence length of few tens of pc. This implies that in general the deflections induced by the Galactic fields upon the CRs reaching us from extragalactic space are dominated by the contribution from the  regular field, even if the root mean square amplitude of the random field could exceed the amplitude of the regular field. Moreover, since we will consider in this work scenarios in which, for a source at 4\,Mpc, the extragalactic turbulent magnetic fields already smear the AD of 10\,EeV protons by more than $10^\circ$, the turbulent Galactic component will provide a subdominant deflection, and hence it will be ignored. 

The regular Galactic magnetic field component is still quite uncertain, with different existing models leading to different predictions for the deflections (see e.g. \cite{su10,ps11,jf12, fa19, un19}. These can be in the ballpark of 10$^\circ$ to 30$^\circ$ for 10\,EeV protons and depend on the direction considered.  For definiteness, we will adopt here the model of Jansson and Farrar (JF12) \cite{jf12}, which was obtained by fitting rotation measures of extragalactic sources and synchrotron polarization data. In this model, the regular field has a disk contribution following the spiral structure of the Galaxy, with reversals in direction between different arms. There is also a toroidal (azymuthal)  halo field having different direction, strength and extent in the northern and southern Galactic hemispheres. Also an $X$-shaped halo component is present, which is axisymmetric and poloidal, playing a relevant role in the CR deflections both towards the Galactic center and outside the disk of the Galaxy. 

The CR deflections in the regular field are obtained by integrating the Lorentz force along the trajectories, leading to
\begin{equation}
    \delta\simeq 10^\circ\frac{10\,{\rm EeV}}{E/Z}\left|\int_0^L \frac{{\rm d}\vec x}{\rm kpc}\times \frac{\vec B}{2\,\mu{\rm G}}\right|.
\end{equation}

\begin{figure}[t]
    \centering
    \includegraphics[width=0.49\textwidth]{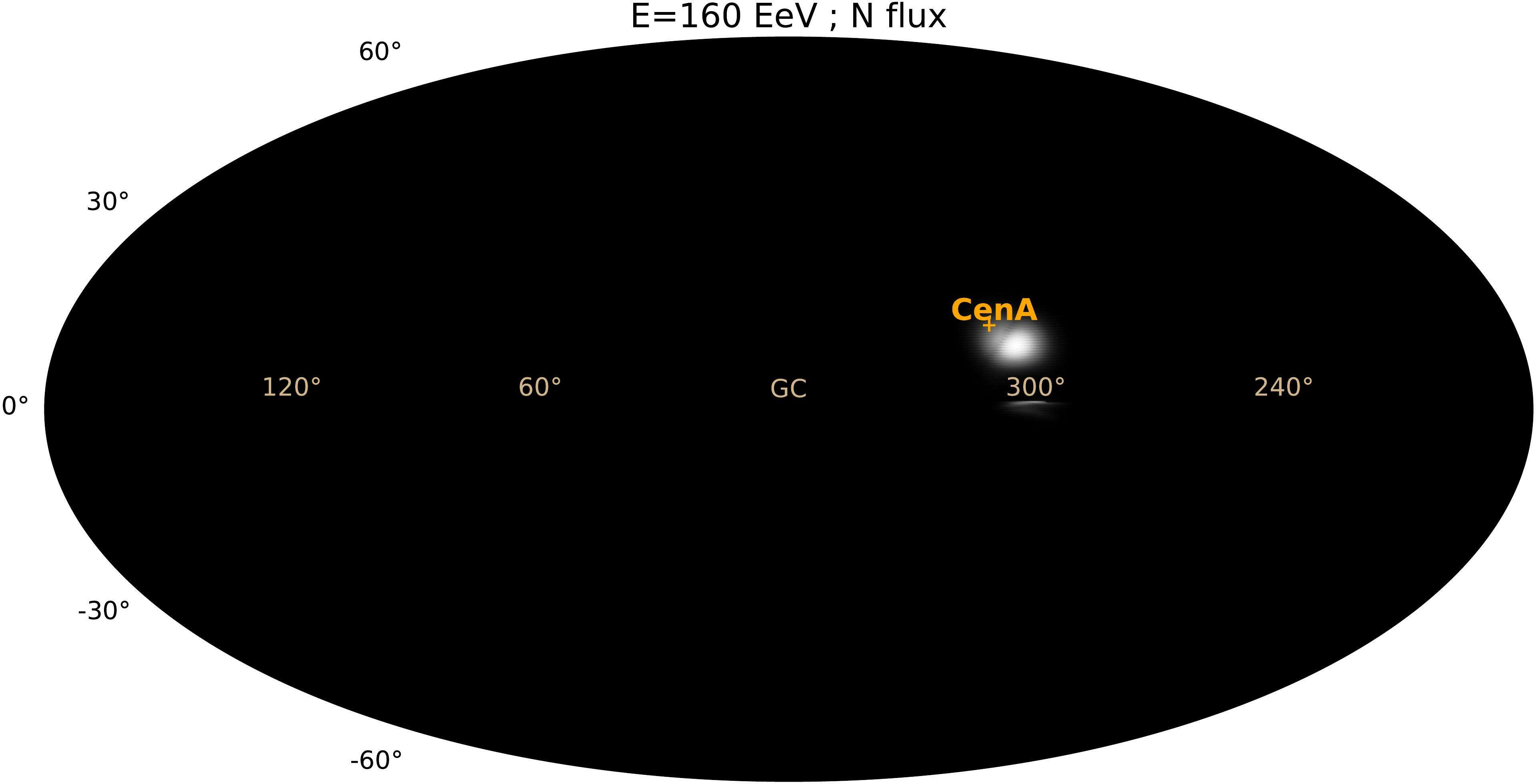} 
    \includegraphics[width=0.49\textwidth]{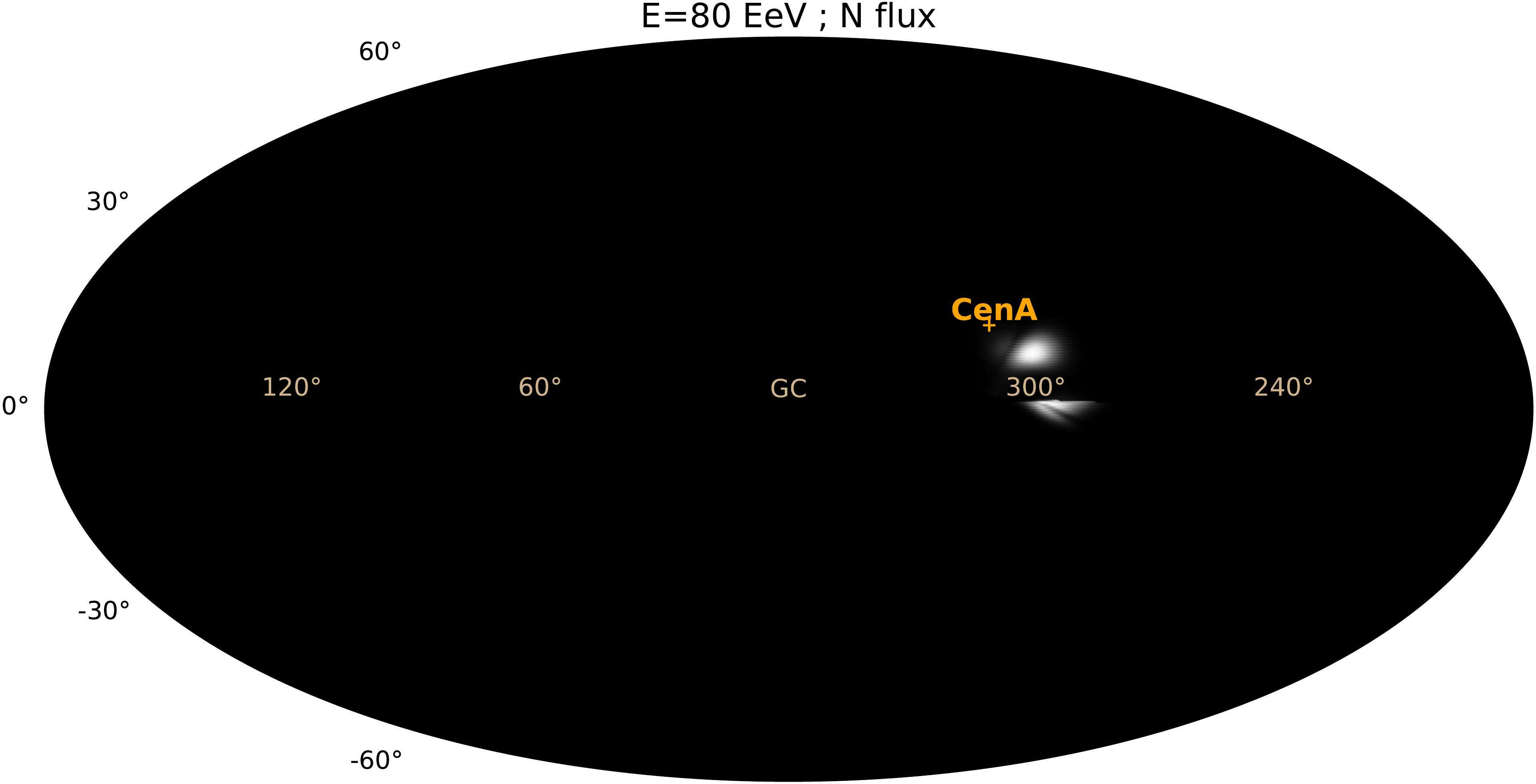} 
    
    \includegraphics[width=0.49\textwidth]{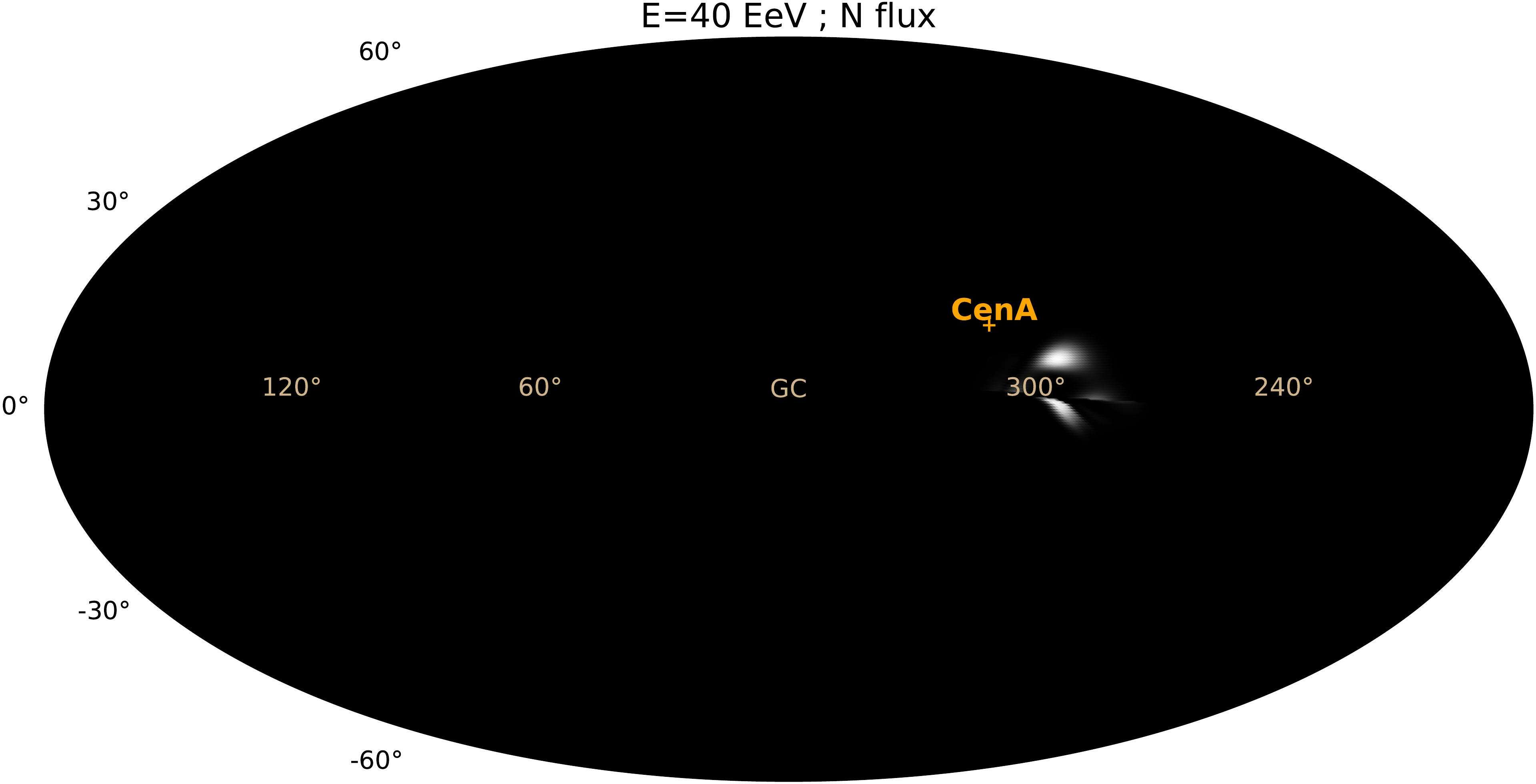} 
    \includegraphics[width=0.49\textwidth]{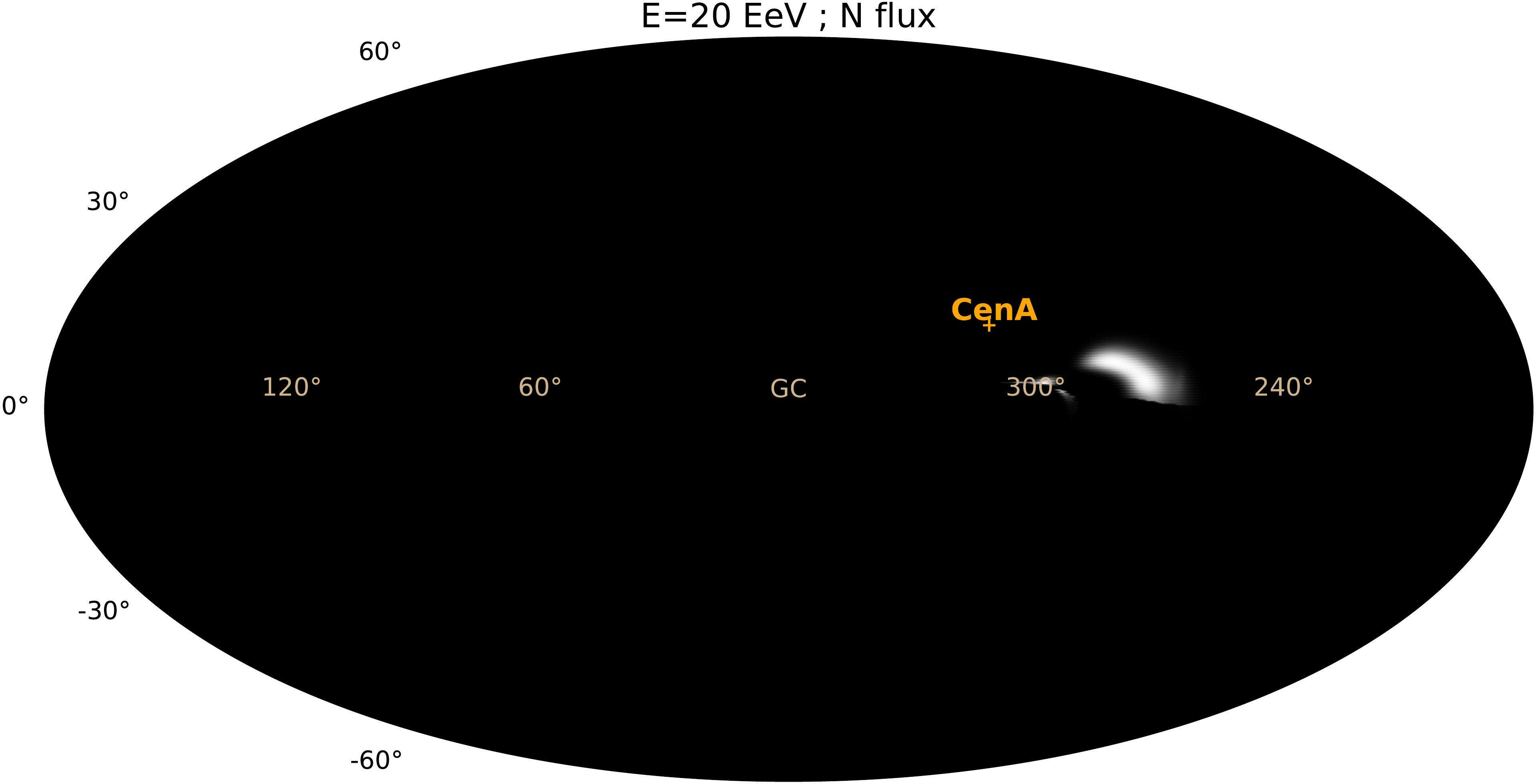}    
    
    \includegraphics[width=0.49\textwidth]{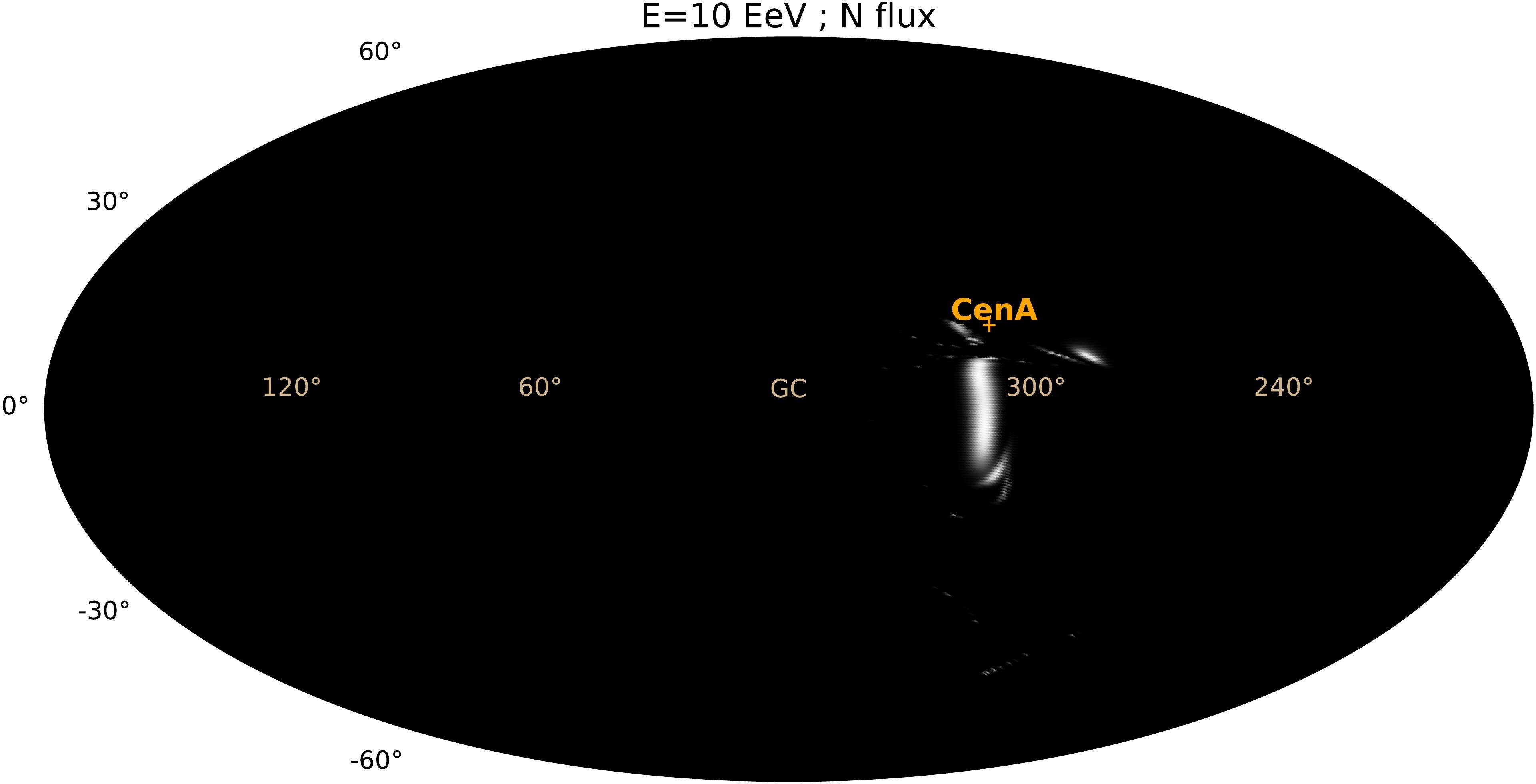} \includegraphics[width=0.49\textwidth]{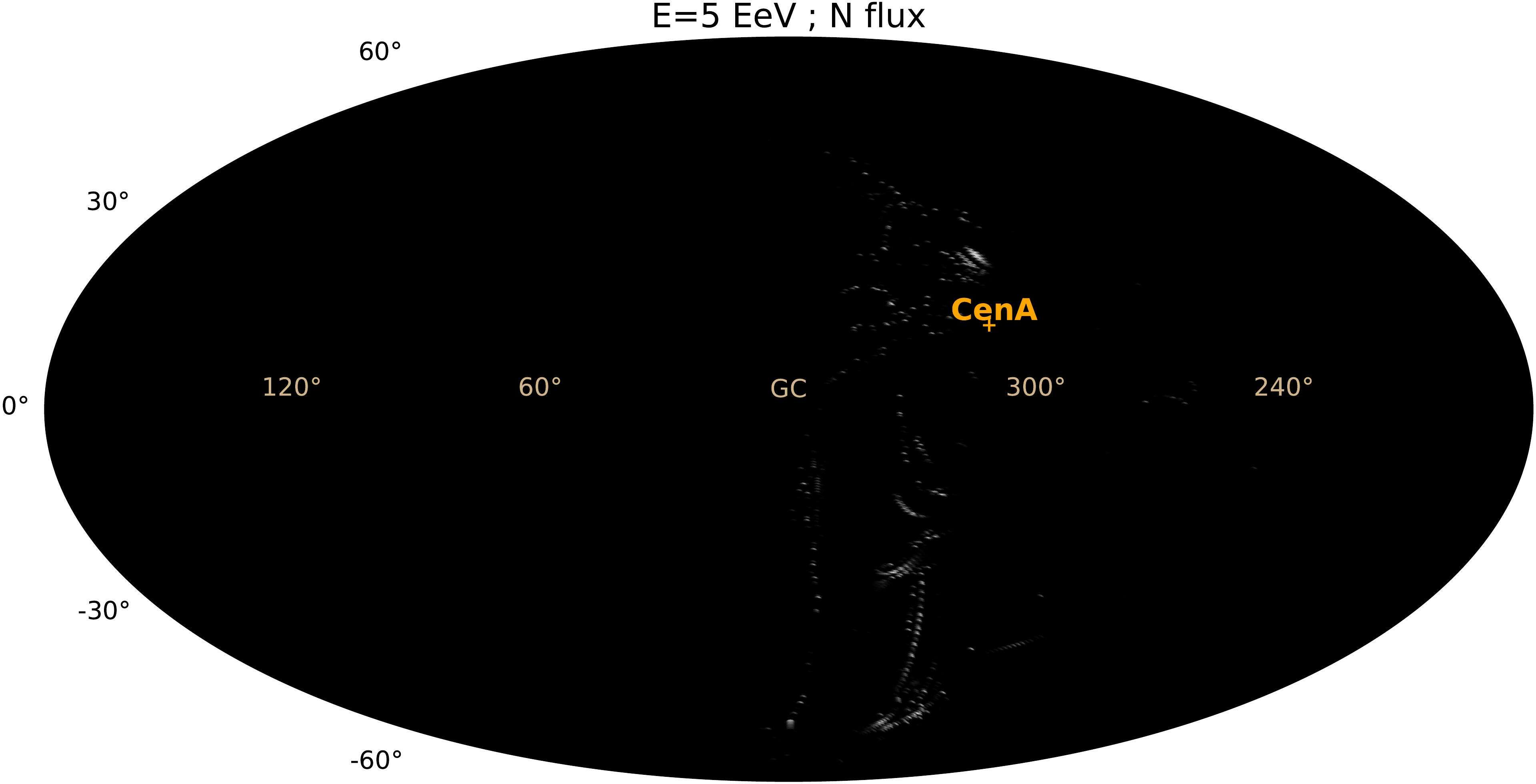}

    \caption{Flux maps of CR nitrogen nuclei emitted by Centaurus~A (orange cross) after being deflected by the Galactic magnetic field, for different energies. We consider here that outside the Galaxy the image of Cen~A is spread with a Gaussian of $5^\circ$ width for all energies.}
    \label{f2}
\end{figure}

In Figure~\ref{f2} we show how the source Centaurus~A, lying at galactic coordinates $(\ell, b)=(309.6^\circ,19.4^\circ$) and at a distance $r_{\rm s}=3.8$~Mpc, would look like from the Earth for different rigidities, corresponding to those of nitrogen nuclei with energies 5, 10, 20, 40, 80 and 160~EeV. The fluxes are obtained by backtracking the corresponding antiparticles and then assigning to each  direction on Earth the flux distribution in the direction of the particle velocity outside the Galaxy. In  Figure~\ref{f2} we consider that the smearing outside the Galaxy is just a Gaussian distribution of 5$^\circ$ spread, so as to allow to identify the appearance of multiple images of the source and to follow their displacement with energy. One can see for instance that the principal image of Cen~A moves, for decreasing energies, towards the Galactic plane and away from the Galactic center,  many more images appear around this region at the lower energies and in particular some lie below the Galactic plane. Given the Galactic latitude of the source, the effects of the halo fields (both the $X$-shaped and the toroidal one) are relevant in the deflections \cite{ke15,ta19}.

\begin{figure}[t]
    \centering
    \includegraphics[width=0.49\textwidth]{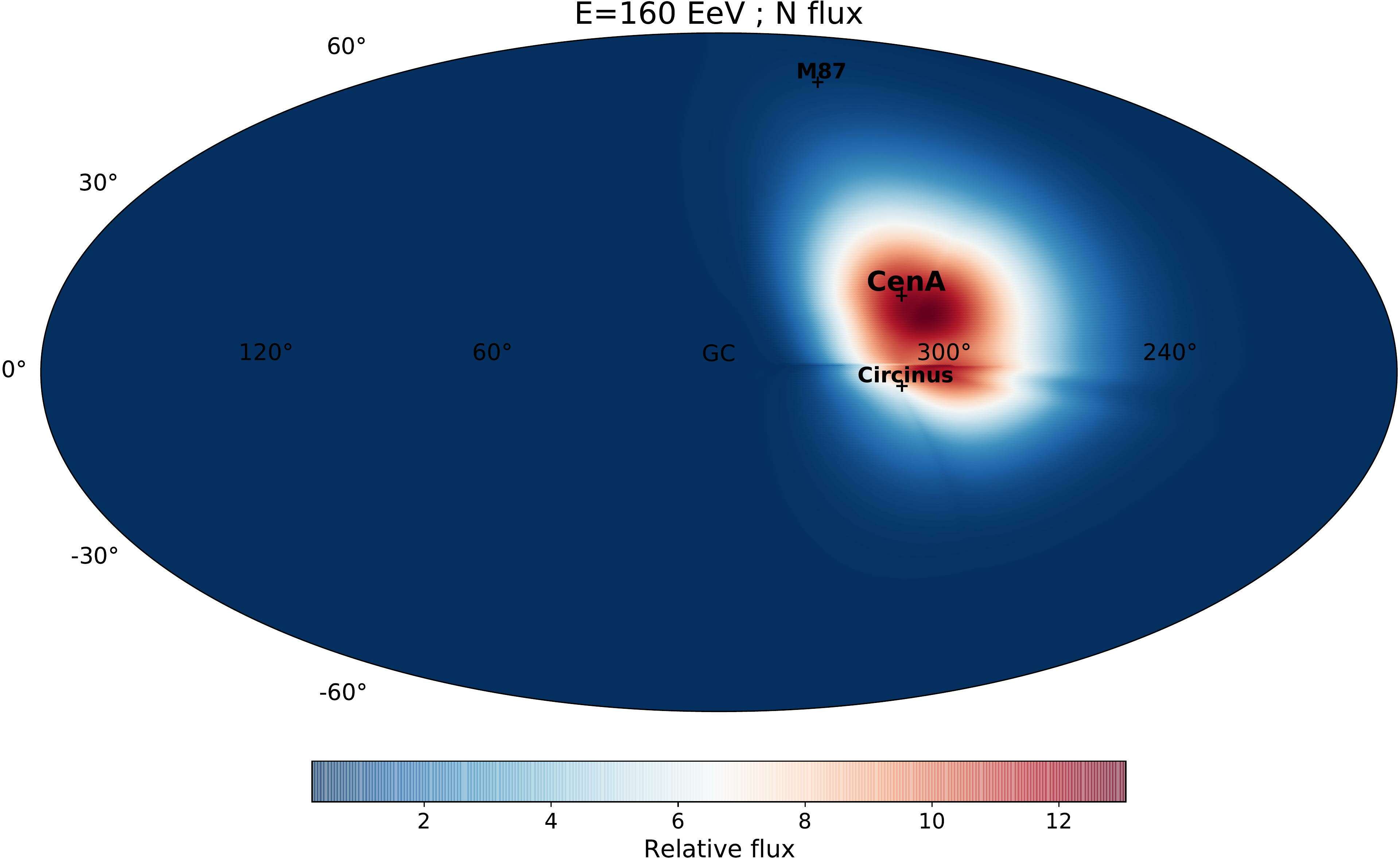} 
    \includegraphics[width=0.49\textwidth]{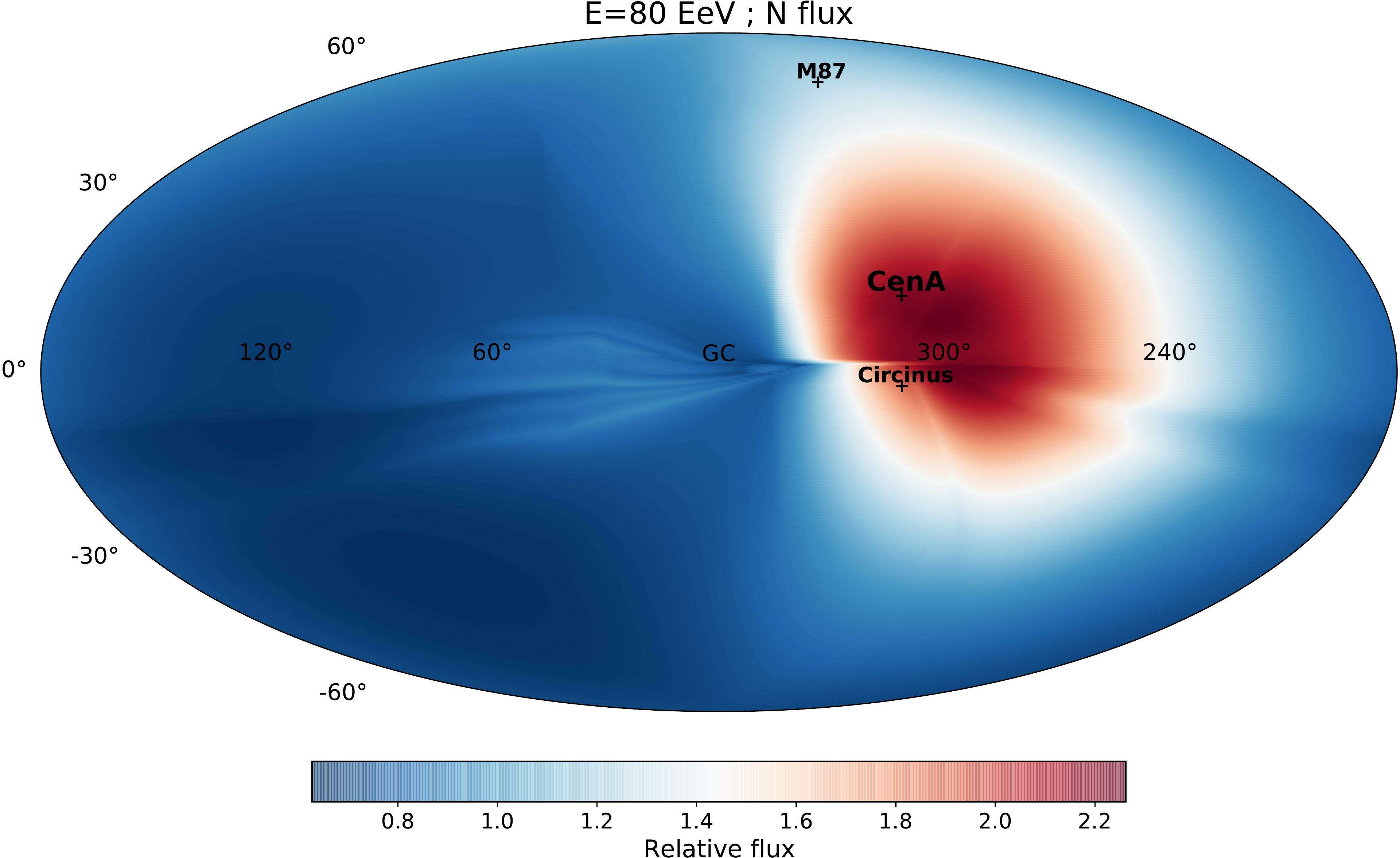} 
    
    \includegraphics[width=0.49\textwidth]{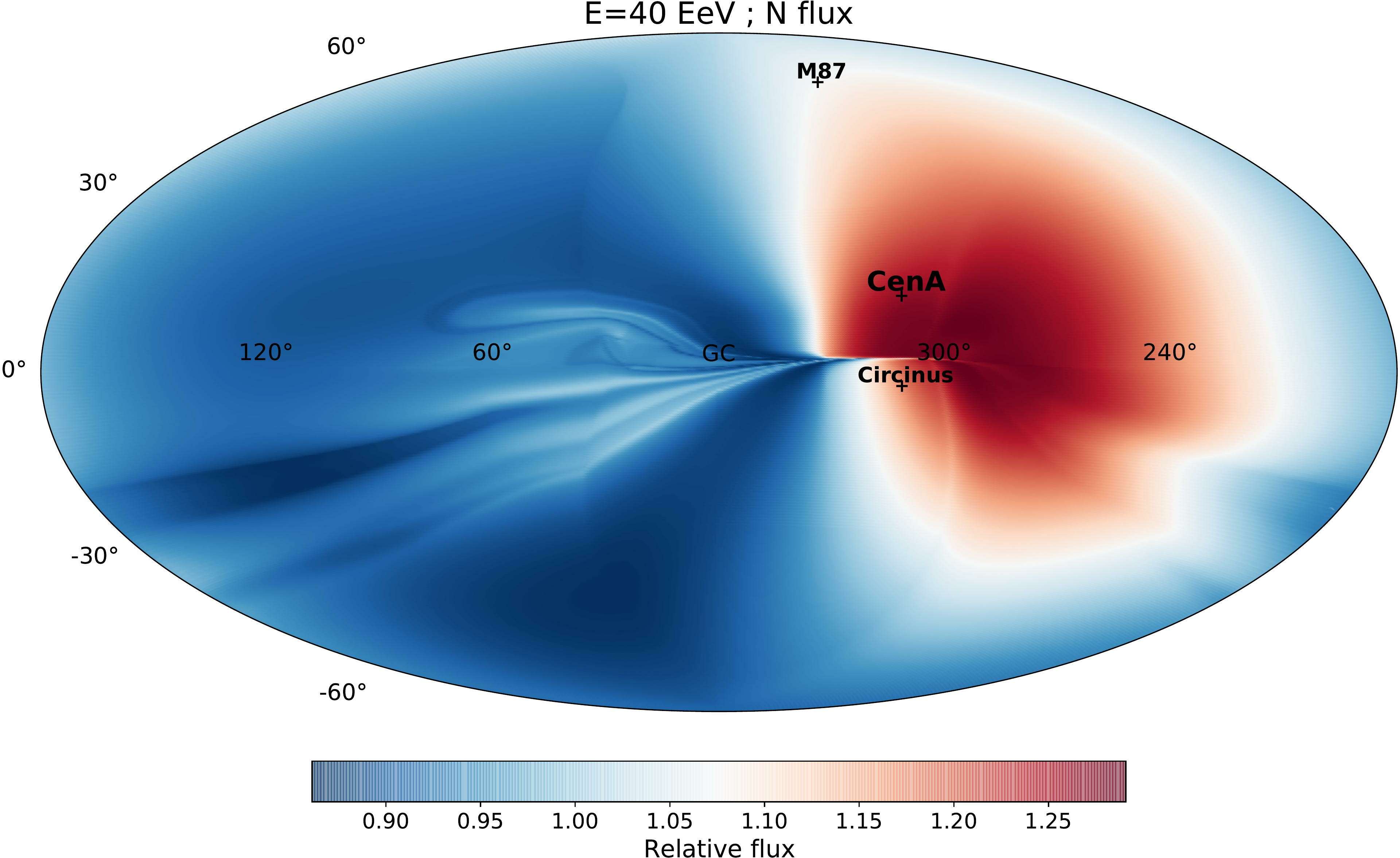} 
    \includegraphics[width=0.49\textwidth]{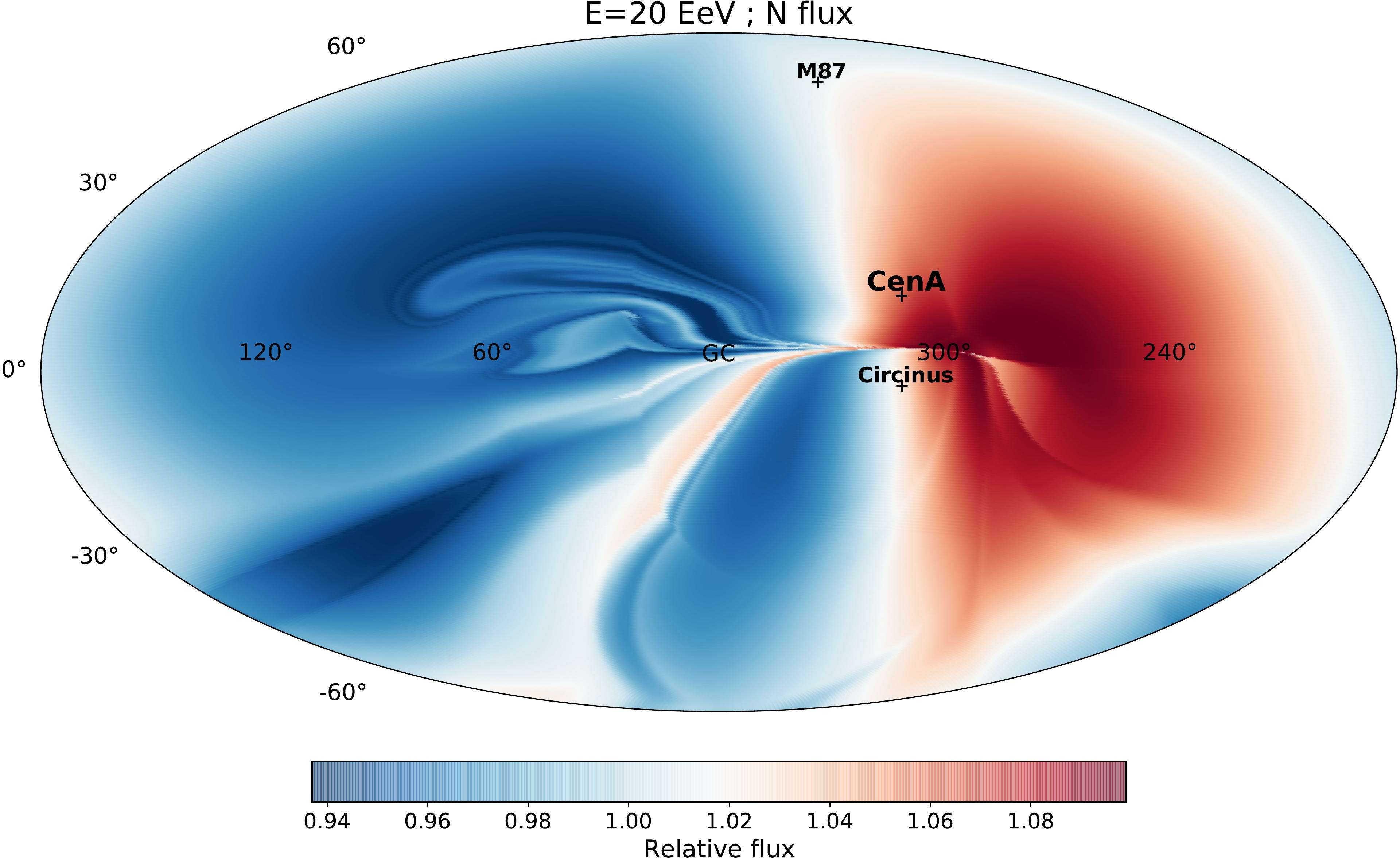}    
    
    \includegraphics[width=0.49\textwidth]{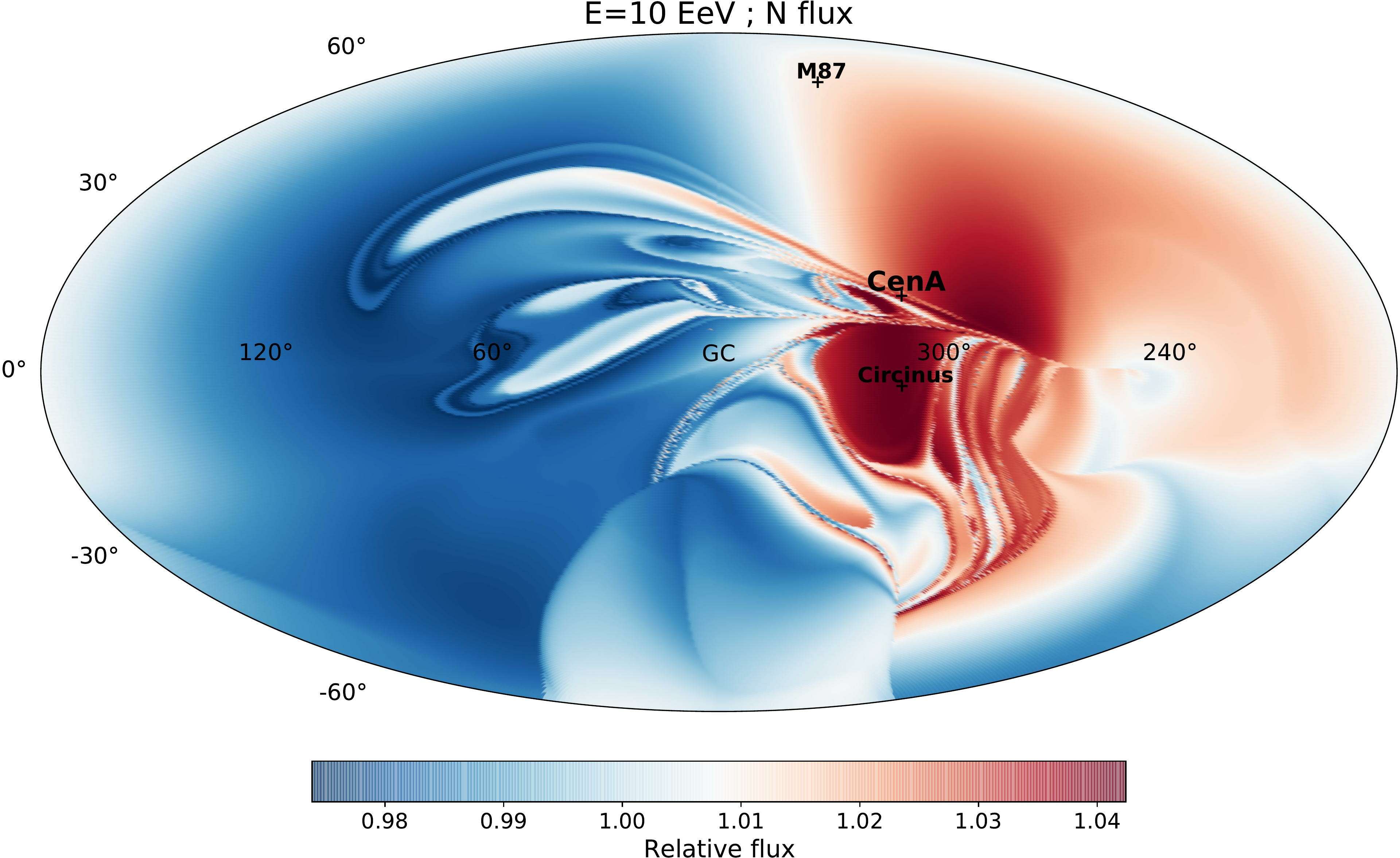} \includegraphics[width=0.49\textwidth]{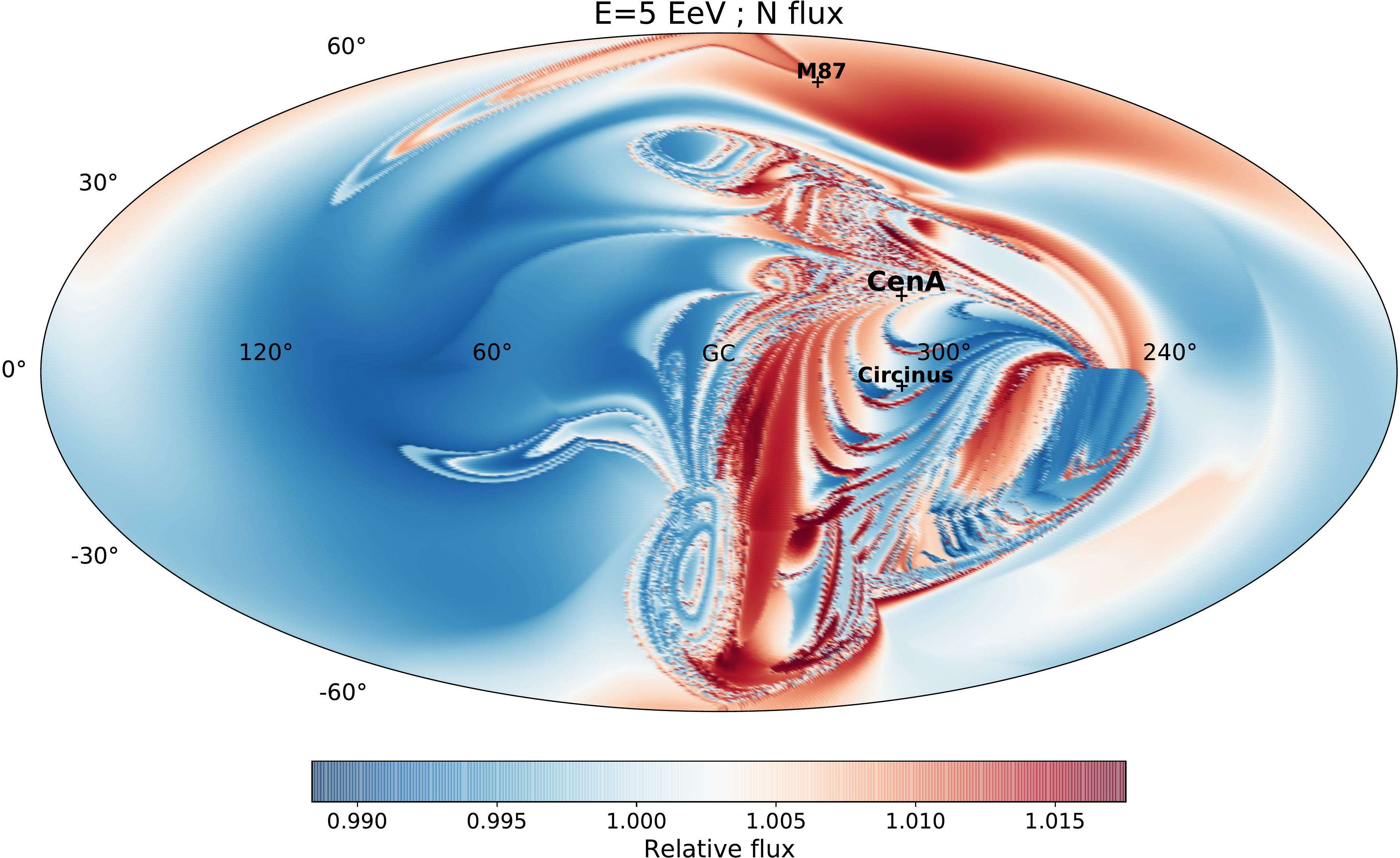}

    \caption{Relative flux maps of CR nitrogen nuclei  emitted by Centaurus~A, after being deflected by the Galactic magnetic field, for different energies. We consider here that outside the Galaxy the CRs have been deflected by a turbulent extragalactic magnetic field of 100~nG and 30~kpc coherence length. Note the different scales of the relative fluxes in each map.}
    \label{f3}
\end{figure}

In Figure~\ref{f3} we consider instead the energy dependent distribution outside the Galaxy which results from the deflections in the turbulent extragalactic field, as was discussed in Section~2, adopting the parameters $B_{\rm rms}=100$~nG and $l_{\rm c}=30$~kpc, as well as taking $ct_i=300$~Mpc. In this case the color code indicates the flux normalized to the average value over all the sky, and one finds that for the lower energies the CRs arrive essentially from all directions, with diminishing contrast. From these figures one can see the relevance of the multiple imaging and flux magnification of the sources \cite{toes}. For a point-like source, the new images appear in pairs when a caustic of the deflection mapping crosses the source location, and for the energy at which they appear these images have divergent magnifications. The spreading of the source image by the turbulent extragalactic field mitigates these divergences, but also implies that the spread-out source could intersect many more caustic lines of the mapping, and more high magnification (critical) lines then appear to the observer, as is apparent in the maps in Figure~\ref{f3}.  One also finds that the principal image may demagnify for decreasing energies, and other secondary images may become brighter than the principal one, giving the impression that the source could be related to a significantly different location in the sky. We also indicated in the maps, as an example, the directions towards some other potential sources, such as the radiogalaxies  M87 and Circinus, near which indeed some extended hot spots appear at some energies, but actually these regions of enhanced fluxes result from the magnetic lensing of the CRs emitted by Cen~A, being then just mirages.

\begin{figure}[t]
    \centering
    \includegraphics[width=0.49\textwidth]{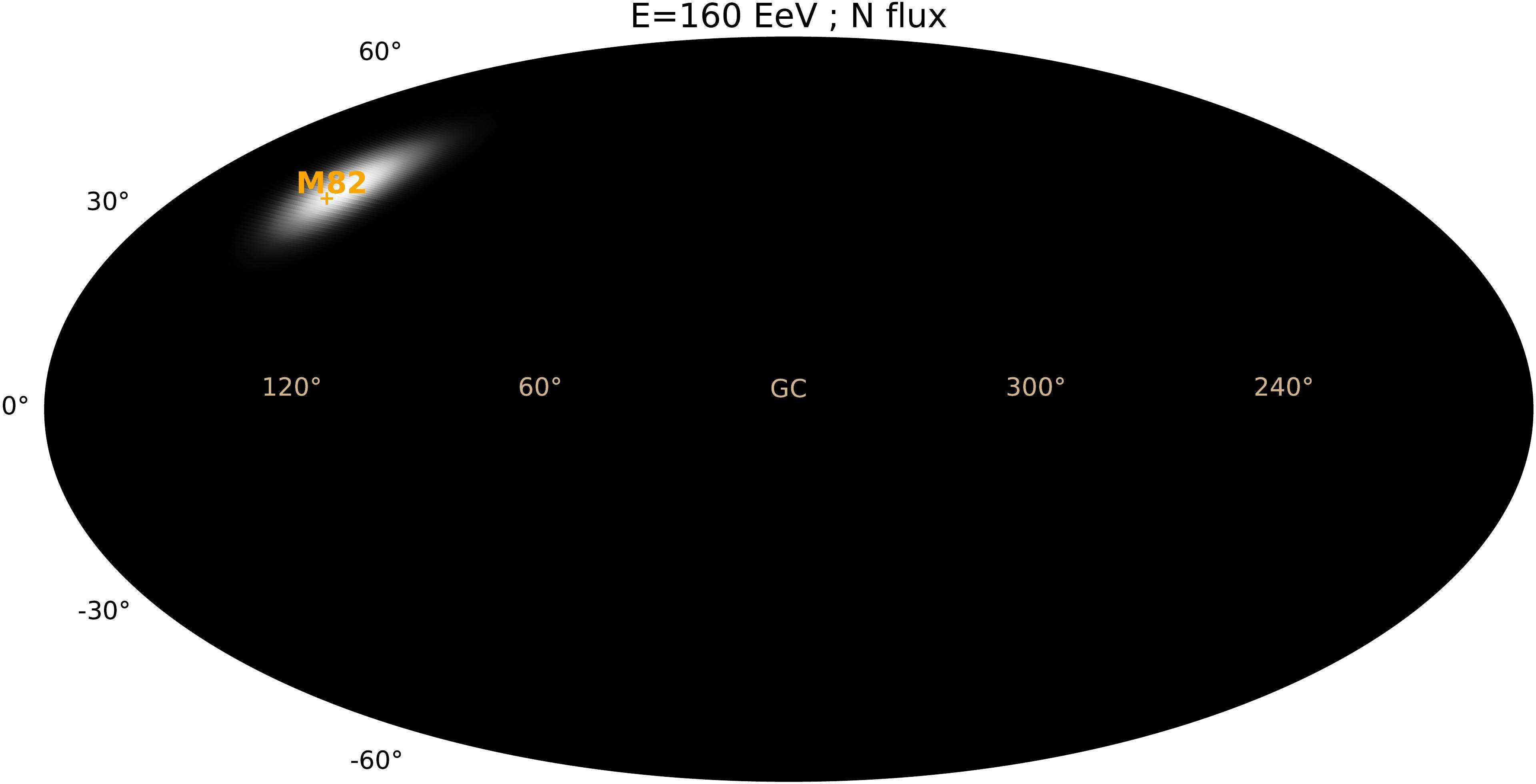} 
    \includegraphics[width=0.49\textwidth]{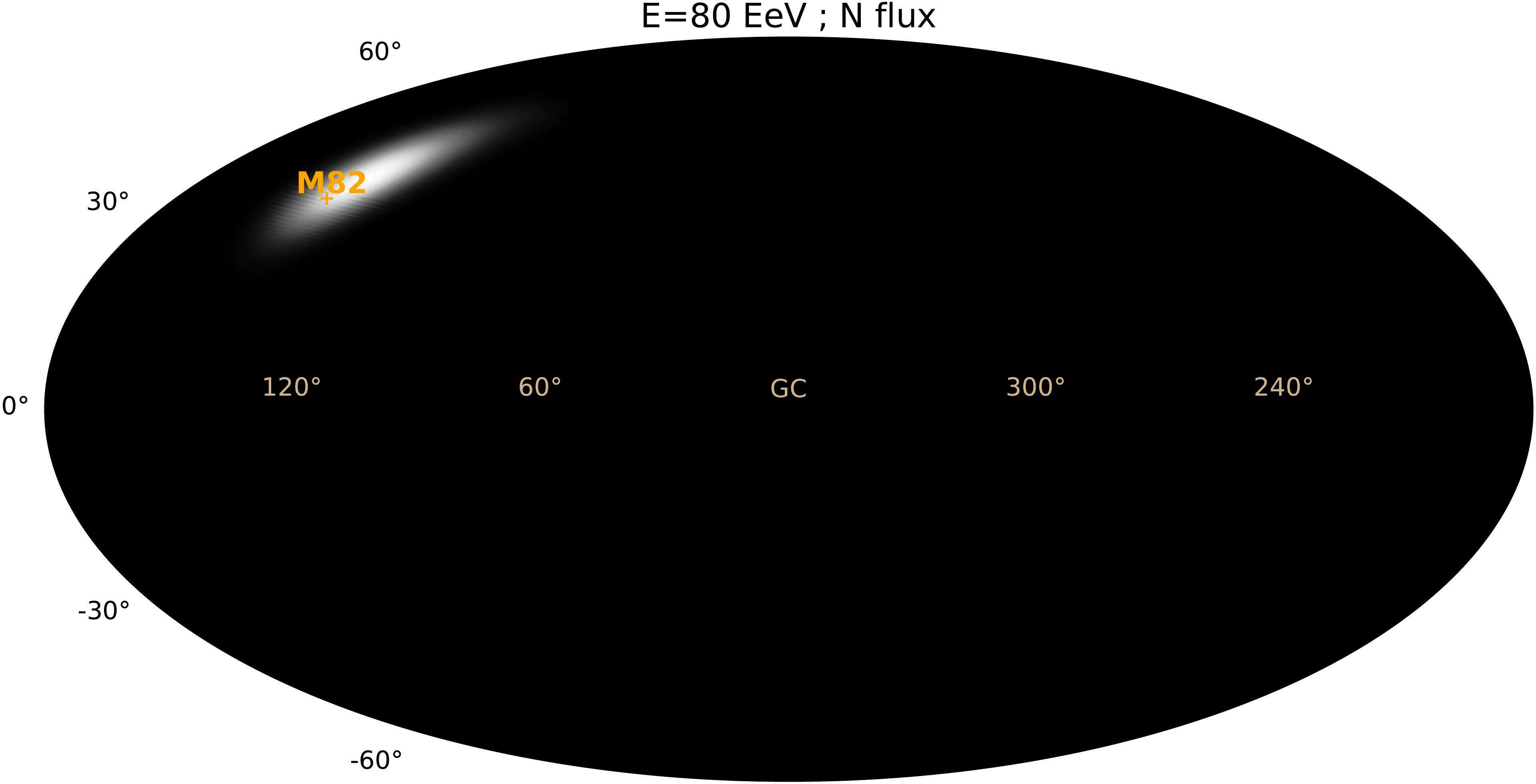} 
    
    \includegraphics[width=0.49\textwidth]{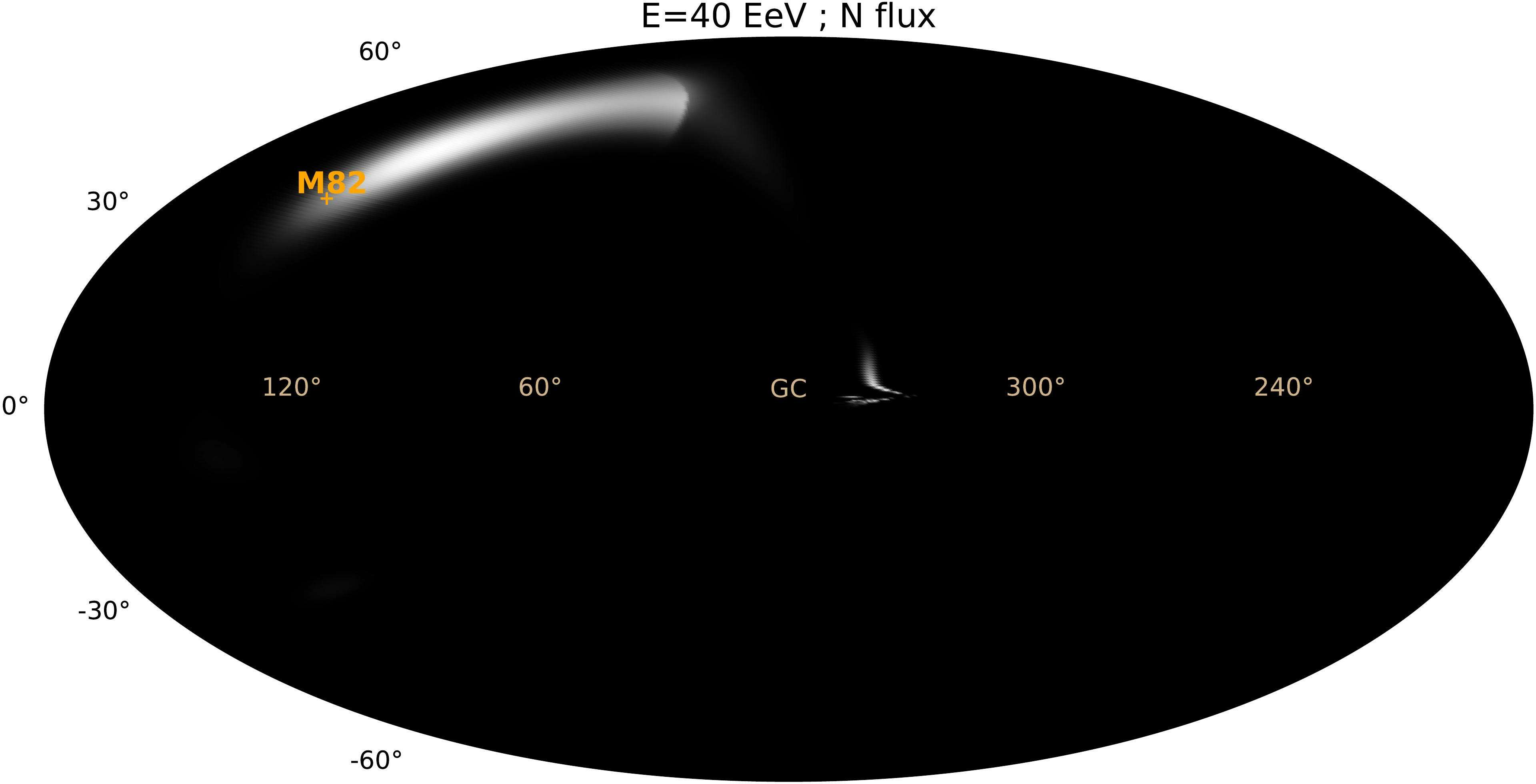} 
    \includegraphics[width=0.49\textwidth]{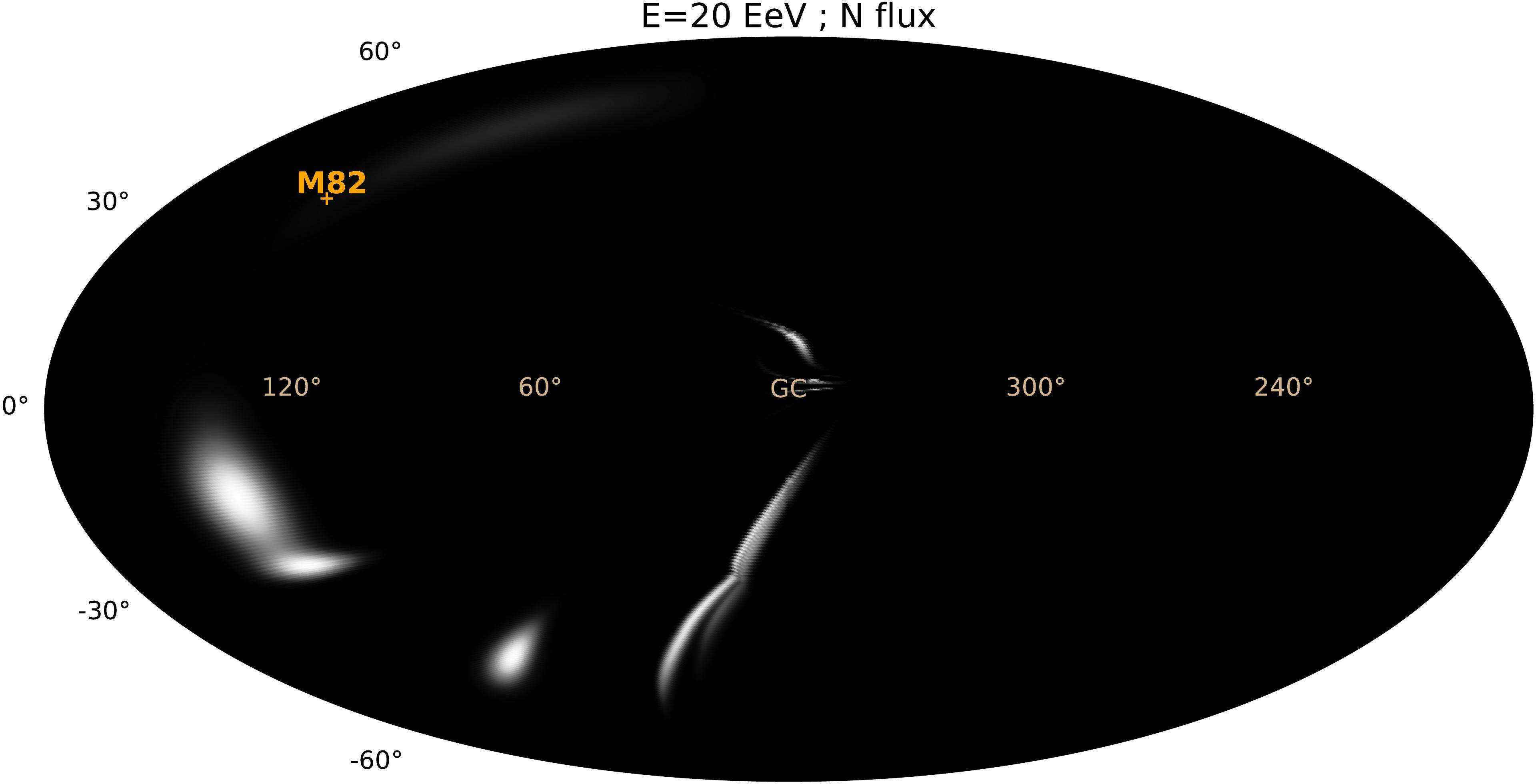}    
    
    \includegraphics[width=0.49\textwidth]{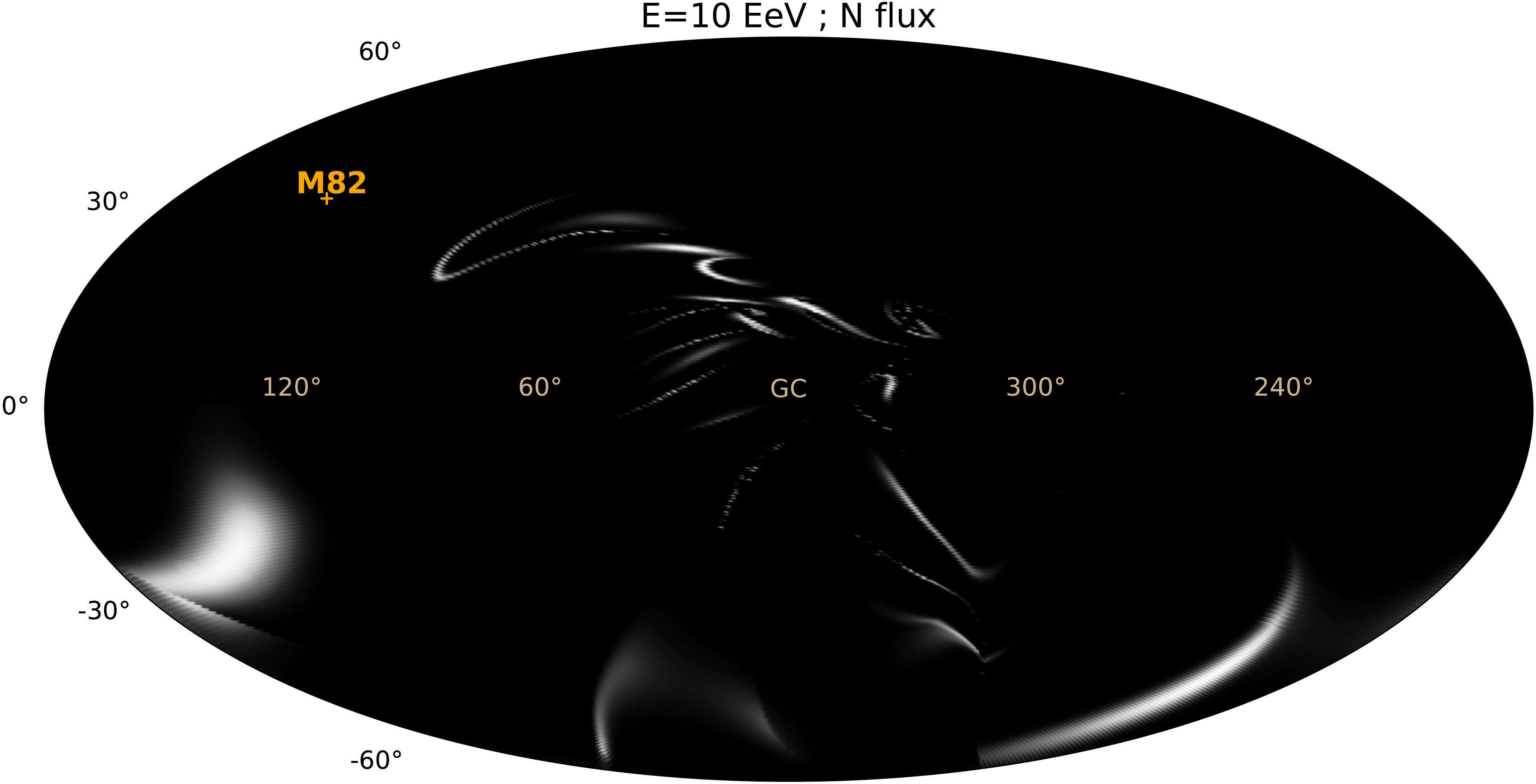} \includegraphics[width=0.49\textwidth]{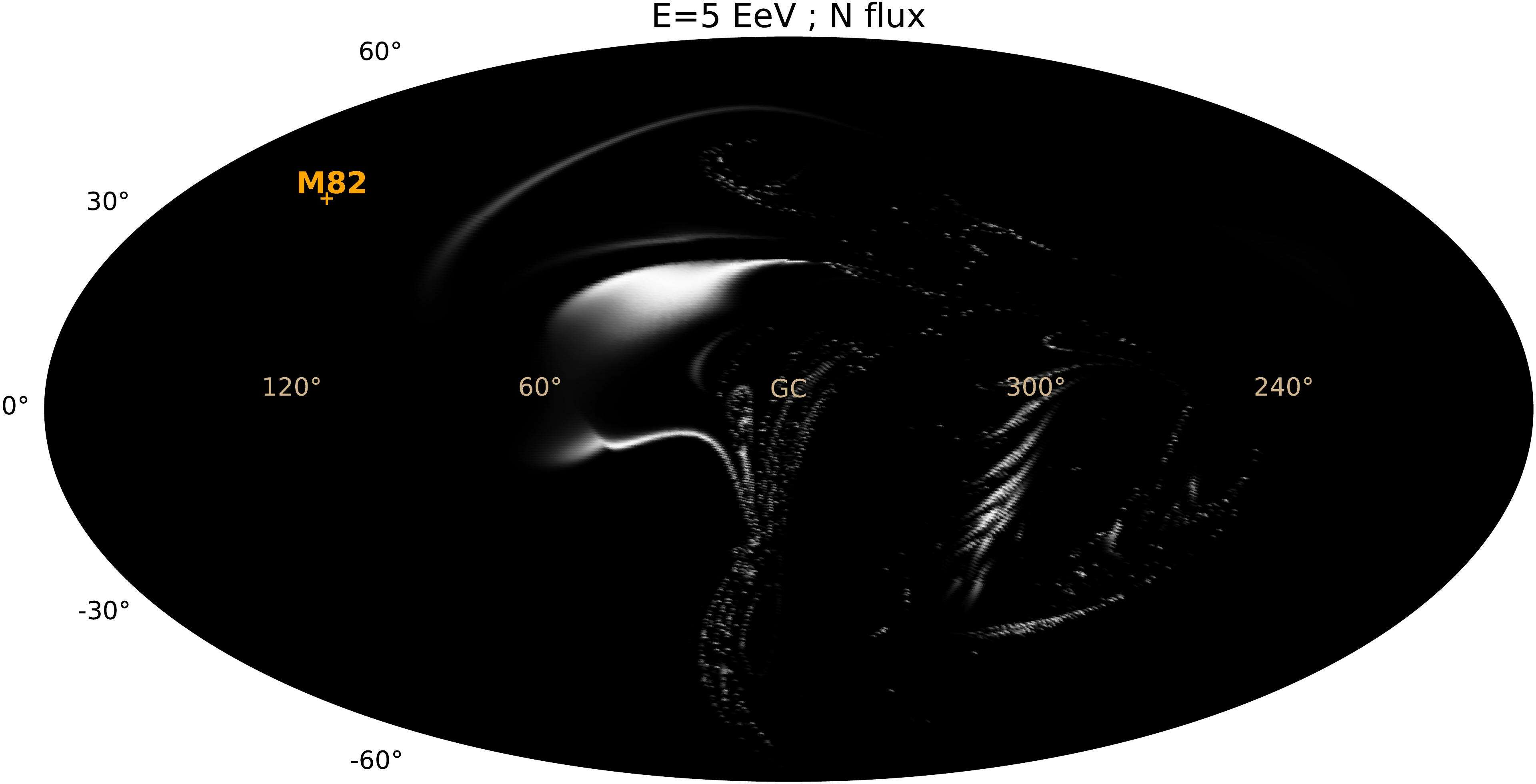}

    \caption{Same as Figure~\ref{f2} but for the source M81/M82.}
    \label{f4}
\end{figure}

\begin{figure}[t]
    \centering
    \includegraphics[width=0.49\textwidth]{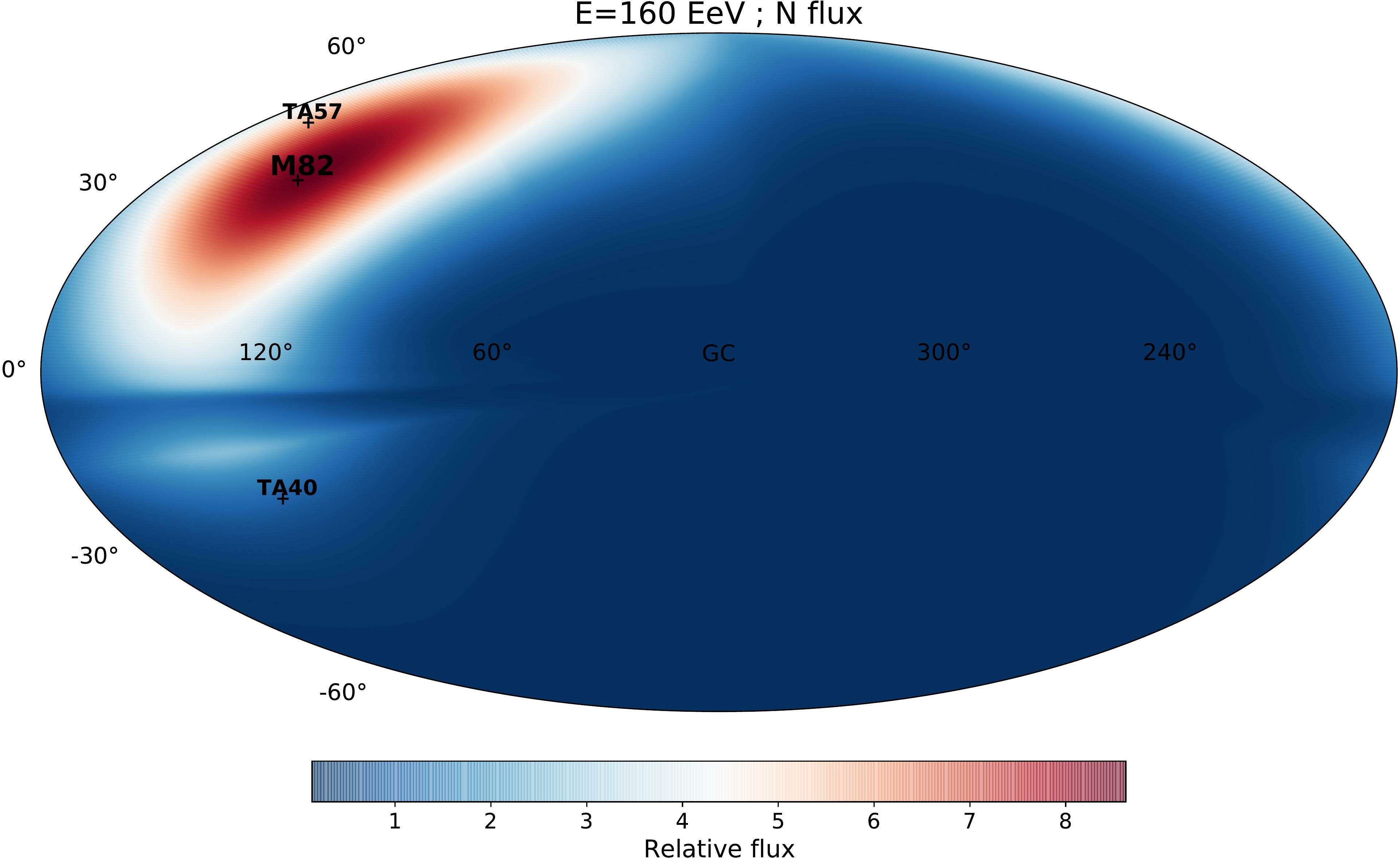} 
    \includegraphics[width=0.49\textwidth]{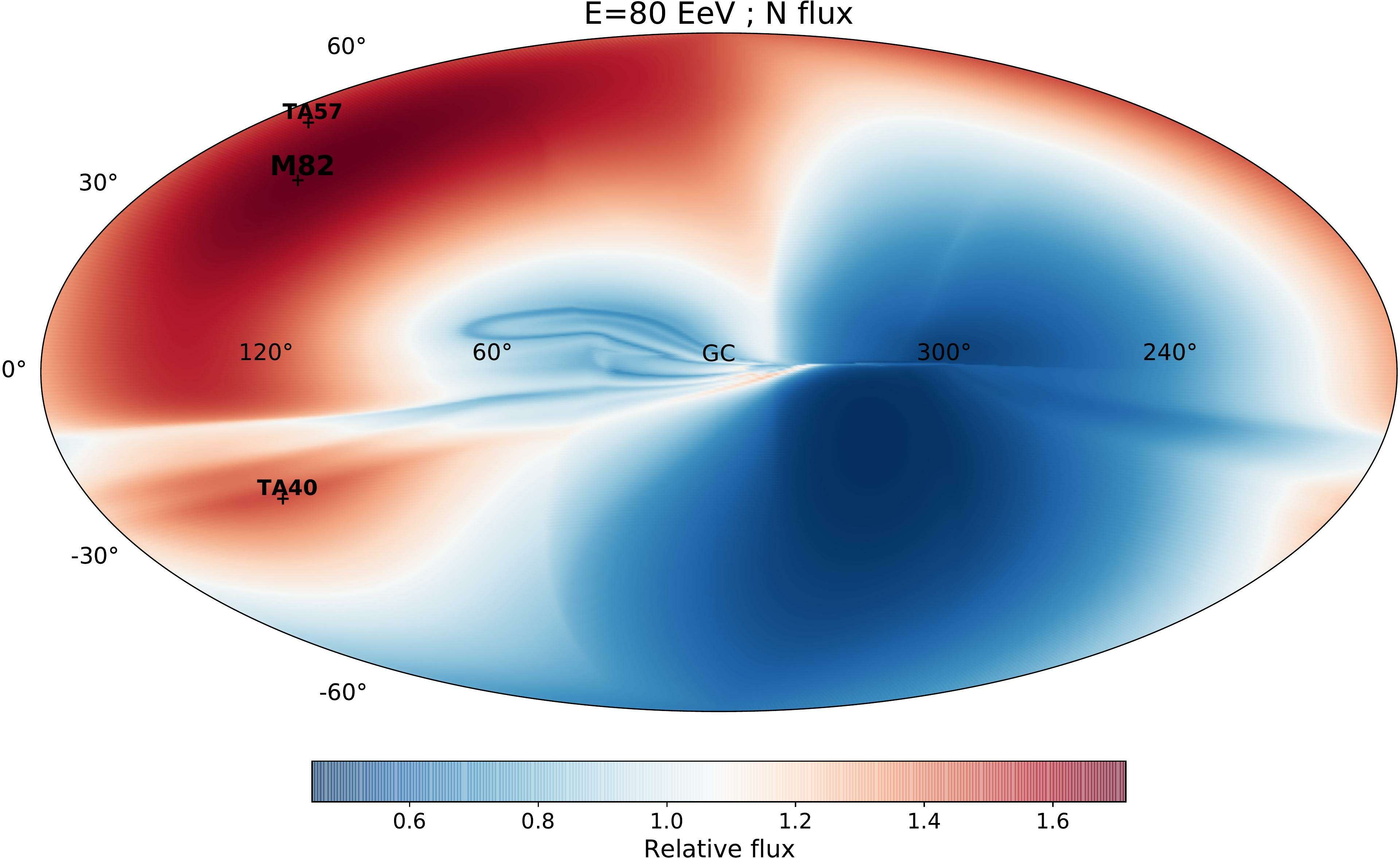} 
    
    \includegraphics[width=0.49\textwidth]{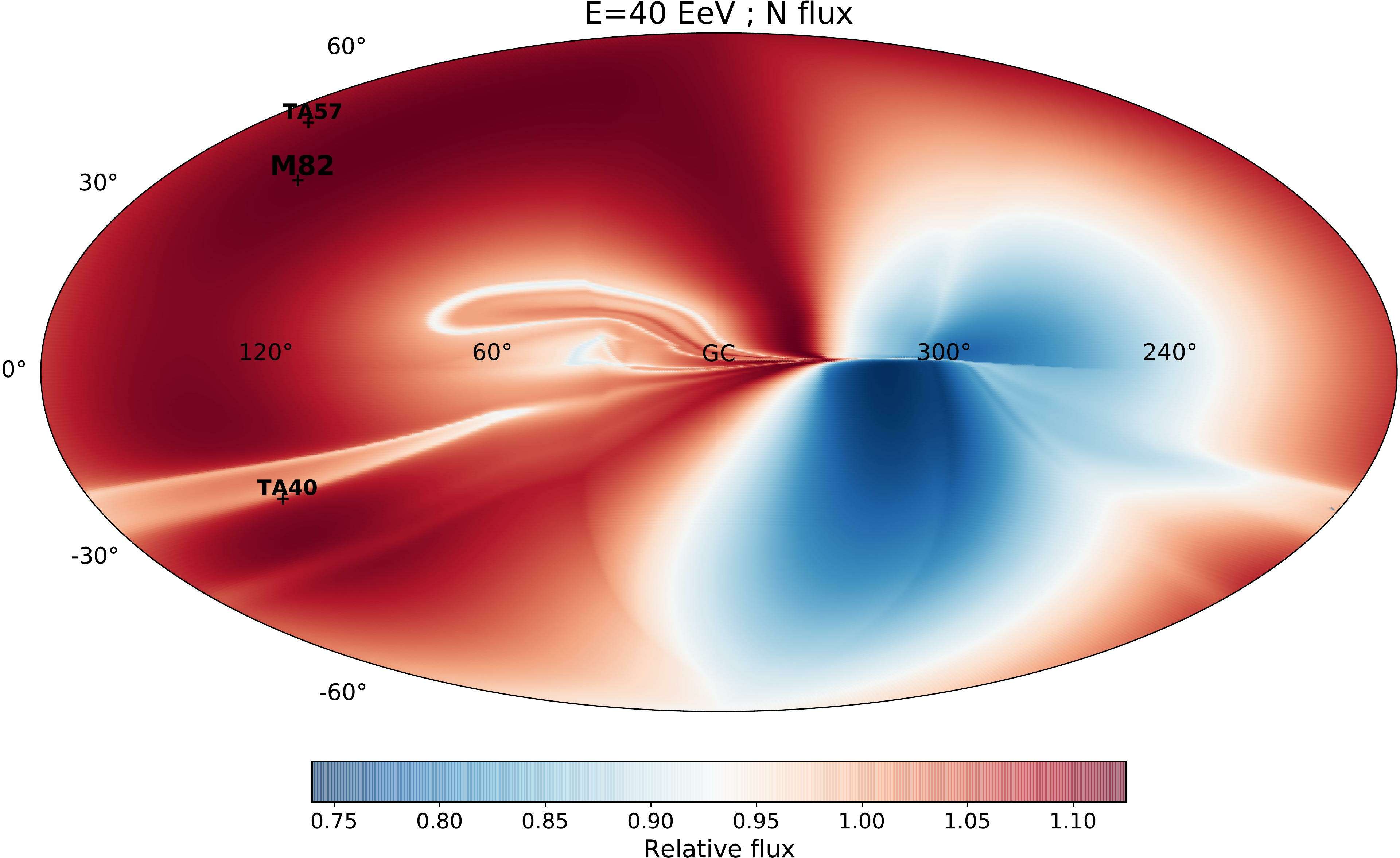} 
    \includegraphics[width=0.49\textwidth]{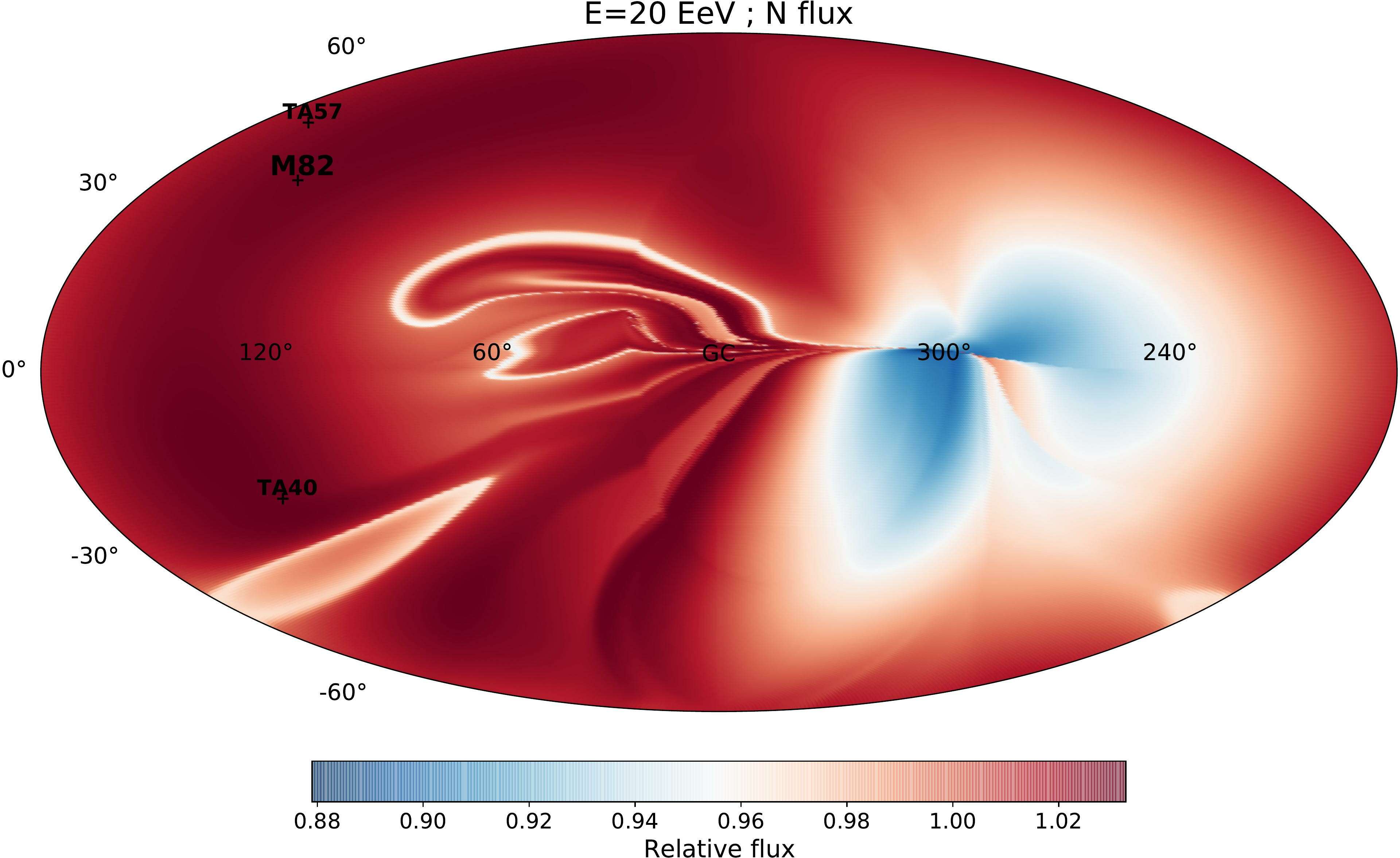}    
    
    \includegraphics[width=0.49\textwidth]{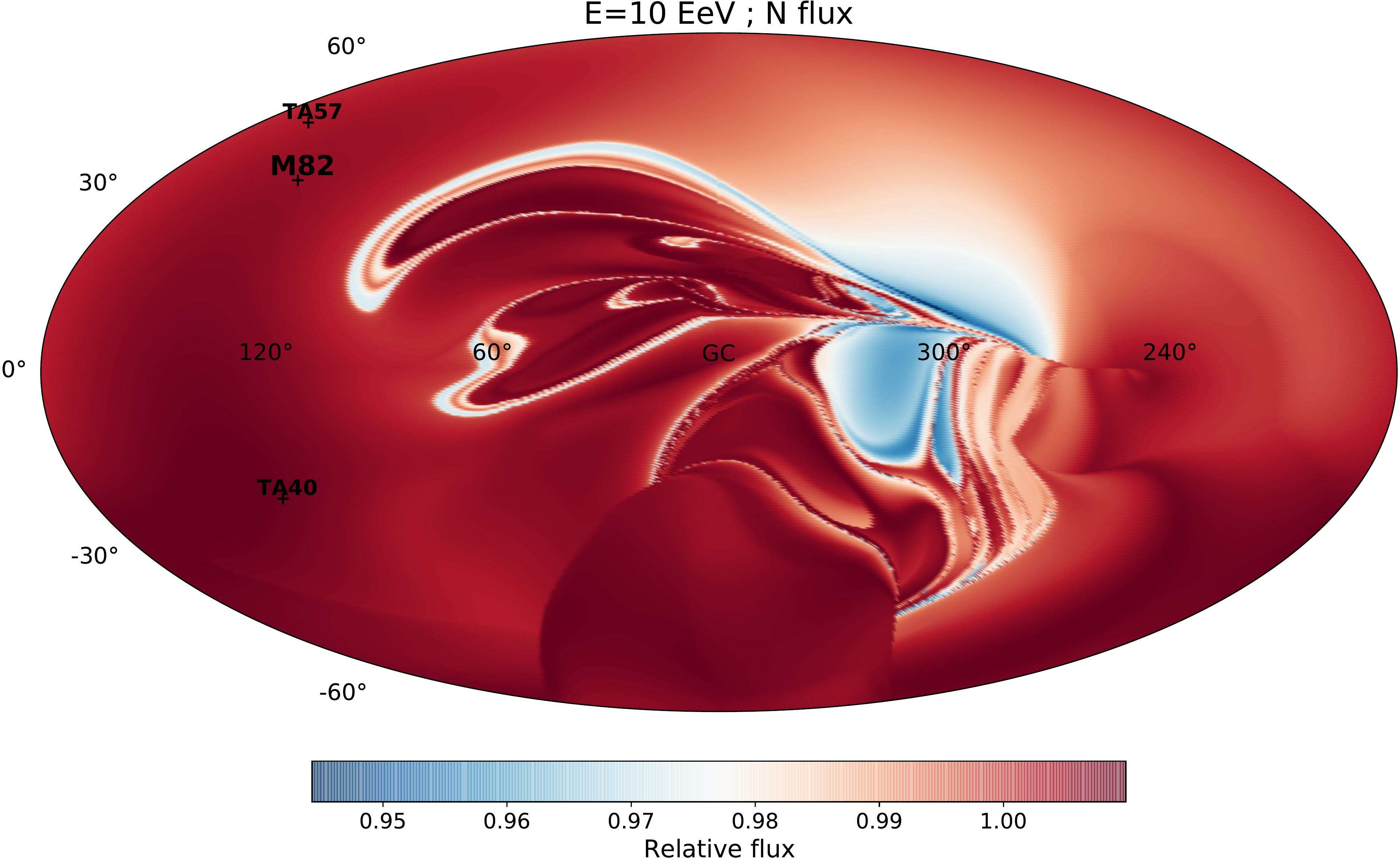} \includegraphics[width=0.49\textwidth]{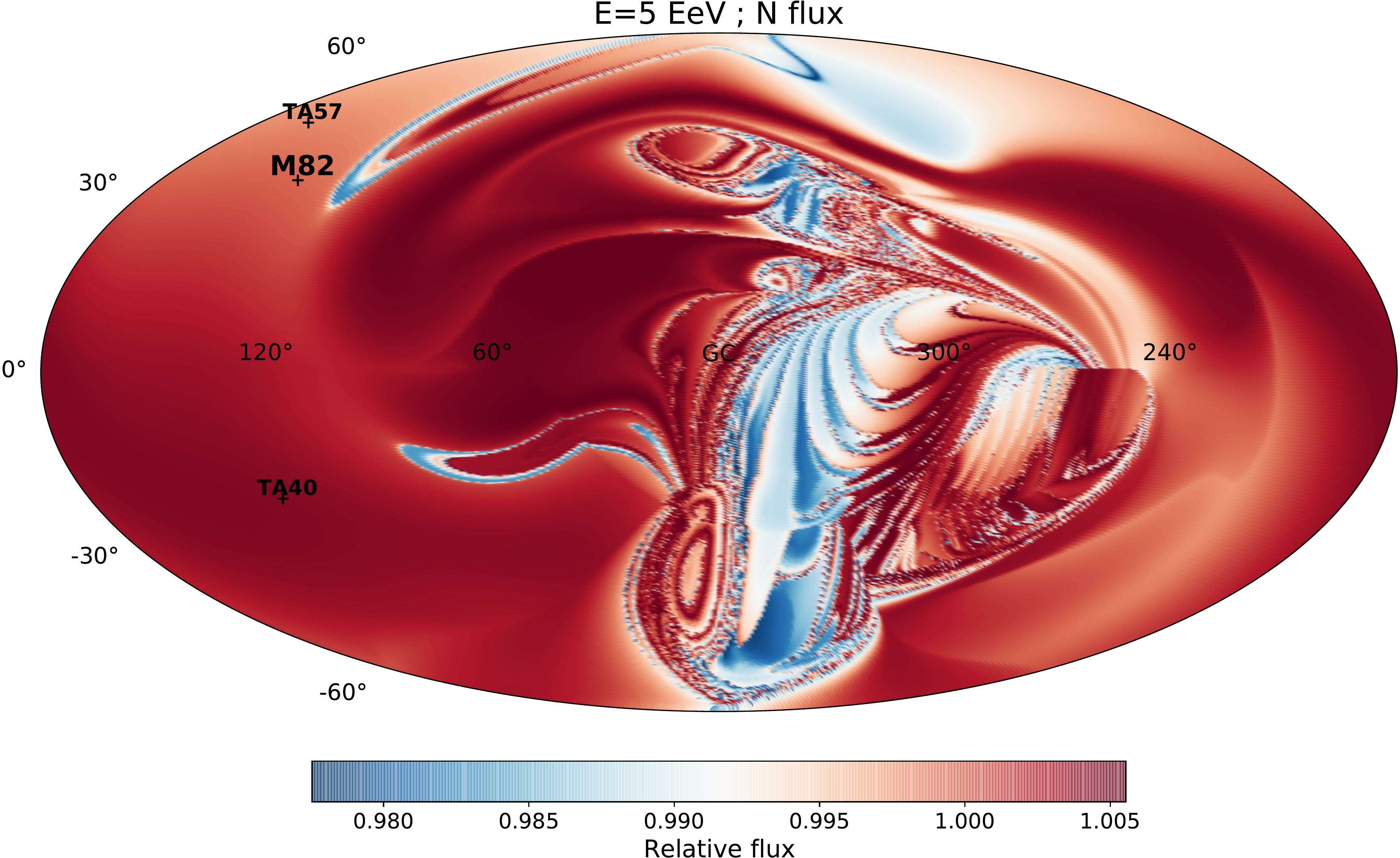}

    \caption{Same as Figure~\ref{f3} but for the source M81/M82.}
    \label{f5}
\end{figure}

Figures~\ref{f4} and \ref{f5} are similar but for the  M81/M82 source, lying at 
$(\ell, b)=(141.4^\circ,40.6^\circ$) and at a distance $r_{\rm s}=3.6$~Mpc. Note that we also indicated in the maps in Figure~\ref{f5} the approximate directions towards the hot spots reported by the Telescope Array Collaboration \cite{tahs} at energies above 57~EeV (TA57) and 40~EeV (TA40).  While the principal image dominating at the highest rigidities is not far from TA57, it is somewhat curious that a bright secondary image of M81/M82 appears at lower energies not far from the direction of TA40.

\section{The Cen~A + M81/M82 mixed composition scenario}

In this section we obtain the arrival direction maps and discuss the main features of the resulting anisotropies for a scenario in which the fluxes above few EeV are dominated by the combined fluxes from the two nearby sources just discussed, Centaurus~A and M81 (or M82). For this we adopt the parameters of the nearby source scenario discussed in ref.~\cite{mo19}, which were shown to lead to very reasonable fits to the observed spectrum and composition, as well as to the overall amplitude of the dipolar anisotropy. Although in that paper just one source was considered, we here include the two sources mentioned, which lie at comparable distances,  adopting for simplicity a similar composition and emission history for both, so that the predicted spectrum and composition ends up being similar.\footnote{We also included, as in \cite{mo19}, an additional extragalactic component, assumed to be isotropic, which becomes relevant below the ankle. This component contributes $\sim 50$\% of the flux at 5\,EeV, and $\sim  20$\% at 10\,EeV, becoming negligible at the higher energies considered.} The extragalactic magnetic field parameters in that scenario are those that were considered in the previous section, i.e. $B_{\rm rms}=100$~nG and $l_{\rm c}=30$~kpc. The sources are assumed to be emitting steadily since  an initial emission time such that $ct_i\simeq 300$~Mpc. The sources are considered to have a power-law spectrum with $\gamma=2.3$ and an exponential cutoff at a rigidity of 10~EV for Cen~A and 20~EV for M81/M82 (given that TA seems to observe a harder spectrum at higher equatorial latitudes, which are closer to the direction towards M81 \cite{tahs}). The adopted fractions for the five representative nuclear species at the sources are $f_{\rm H}=0.45$, $f_{\rm He}=0.09$, $f_{\rm N}=0.31$, $f_{\rm Si}=0.15$ and $f_{\rm Fe}=0$. 

We will neglect for simplicity the interactions during propagation, since we consider relatively low emission redshifts, low cutoff rigidities and  nearby sources. Note that approaching the highest rigidities, where the interactions could become more relevant,  the propagation becomes closer to rectilinear, so that the contribution come mostly from recent emission and hence short pathlengths, while it is only at lower rigidities that the contribution from the earlier emission (and hence larger pathlengths) is relevant. Anyway, some effect from interactions (eventually also from radiation or gas at or around the sources) could affect the heavier components at the highest energies explored, something we will not analyze here.
We will consider for definiteness that at low energies Cen~A provides 80\% of the flux from the nearby sources while M81/M82 provides the remaining 20\%, so that the large scale dipolar distribution at low energies points not far from the Cen~A direction. The relative contributions from the two sources could be in principle used as an extra parameter to adjust the final results (as could be also done with the cutoff energies, relative elemental fractions or initial emission times).

\begin{figure}[t]
    \centering
    \includegraphics[width=0.49\textwidth]{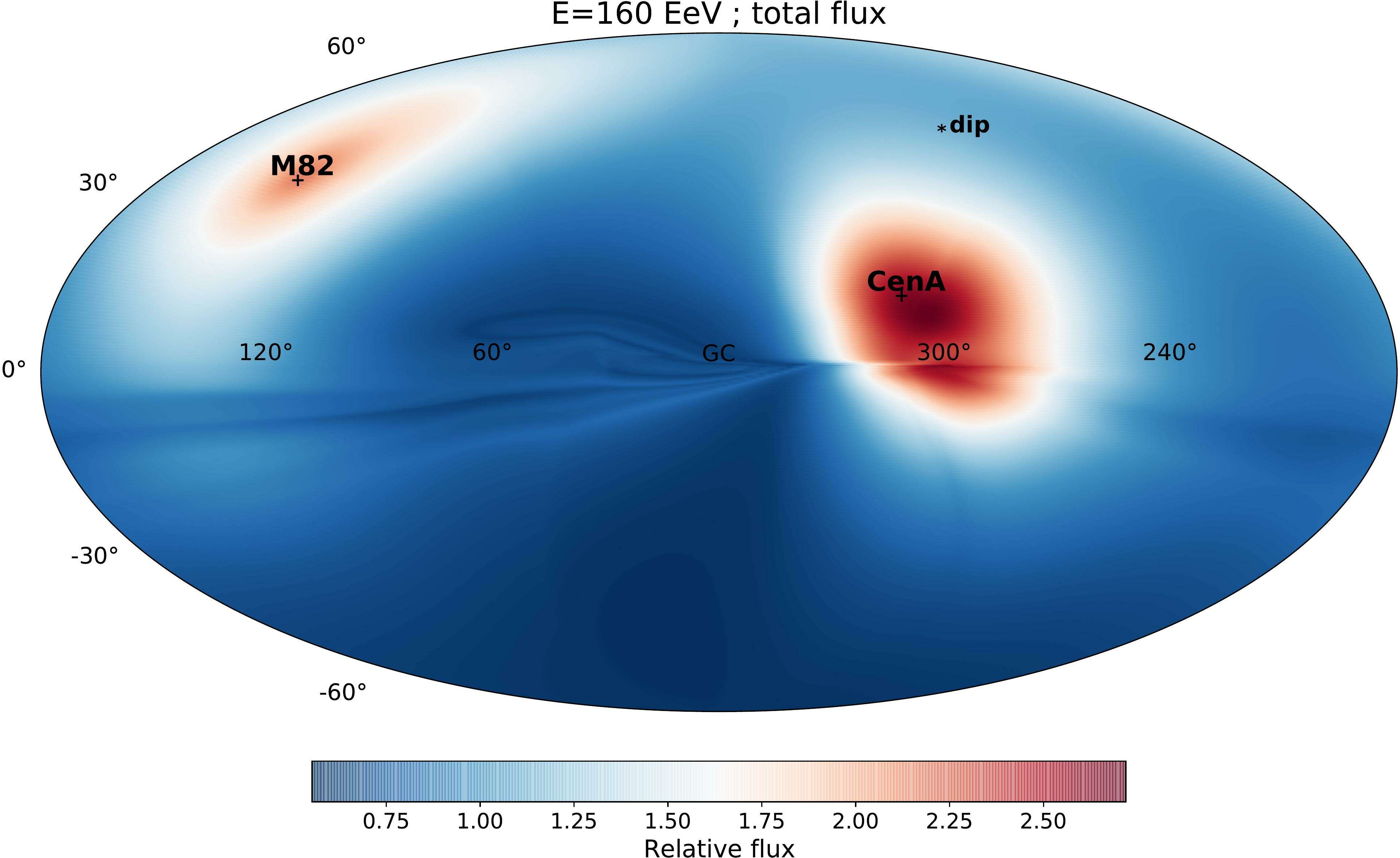} 
    \includegraphics[width=0.49\textwidth]{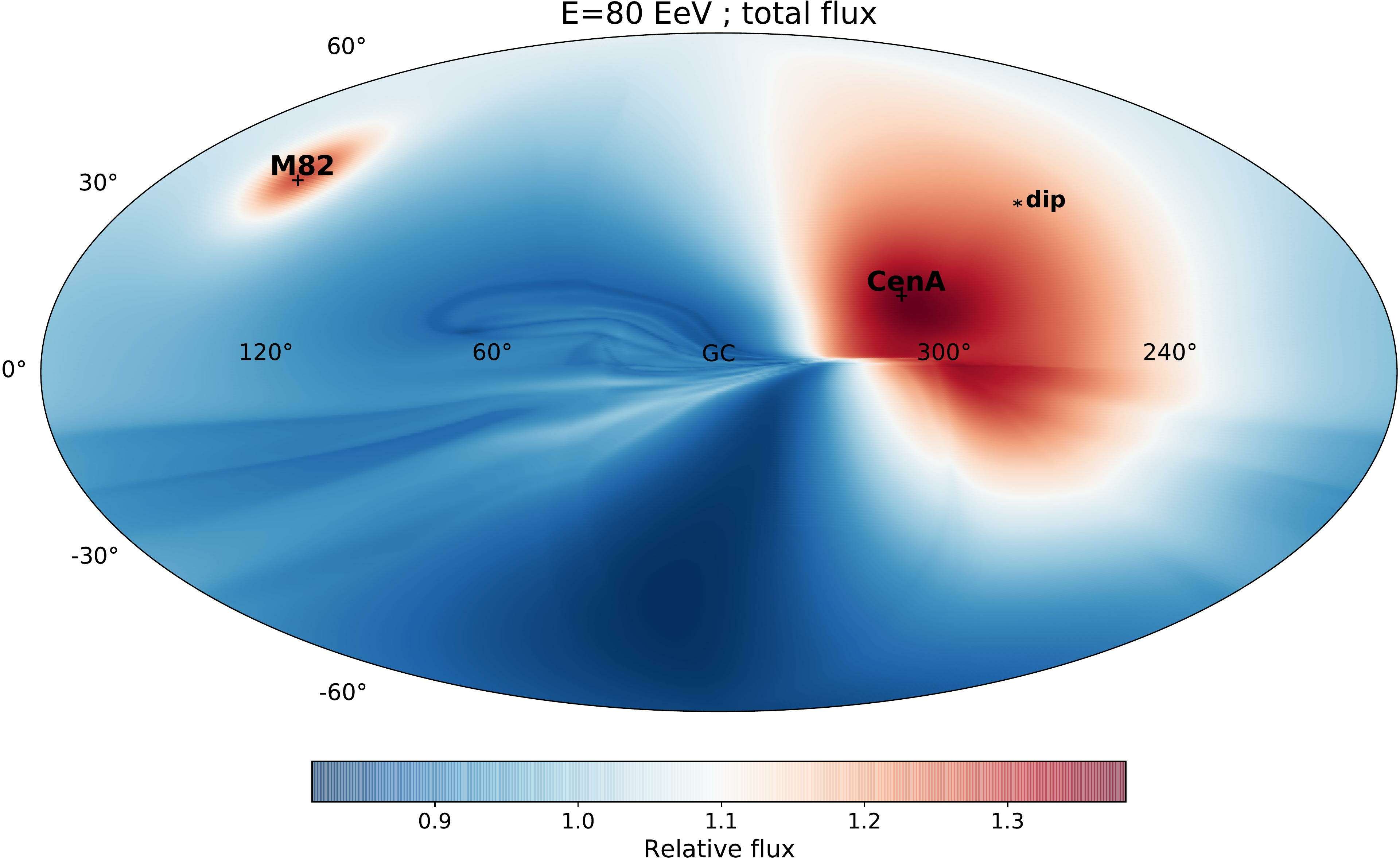} 
    
    \includegraphics[width=0.49\textwidth]{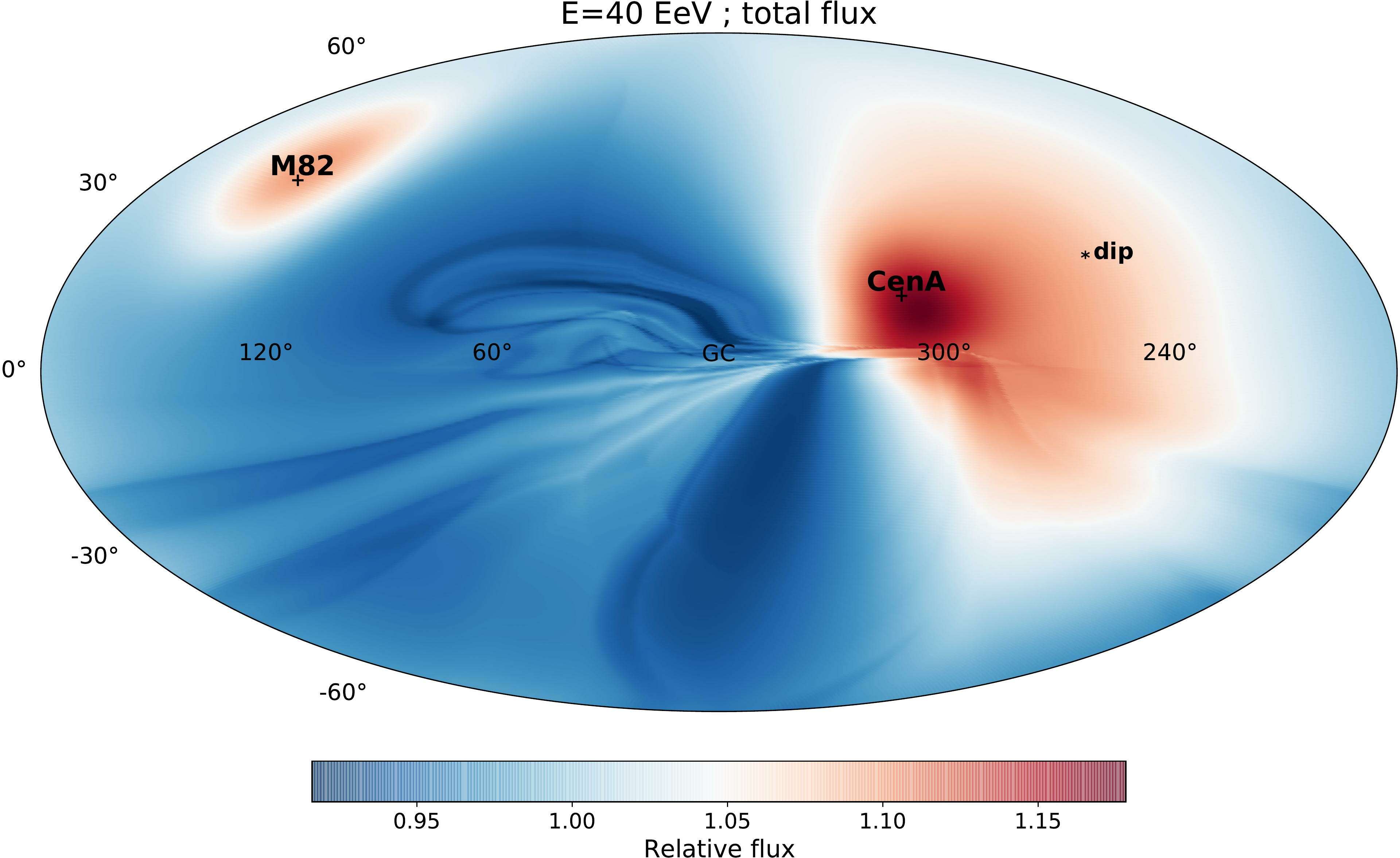} 
    \includegraphics[width=0.49\textwidth]{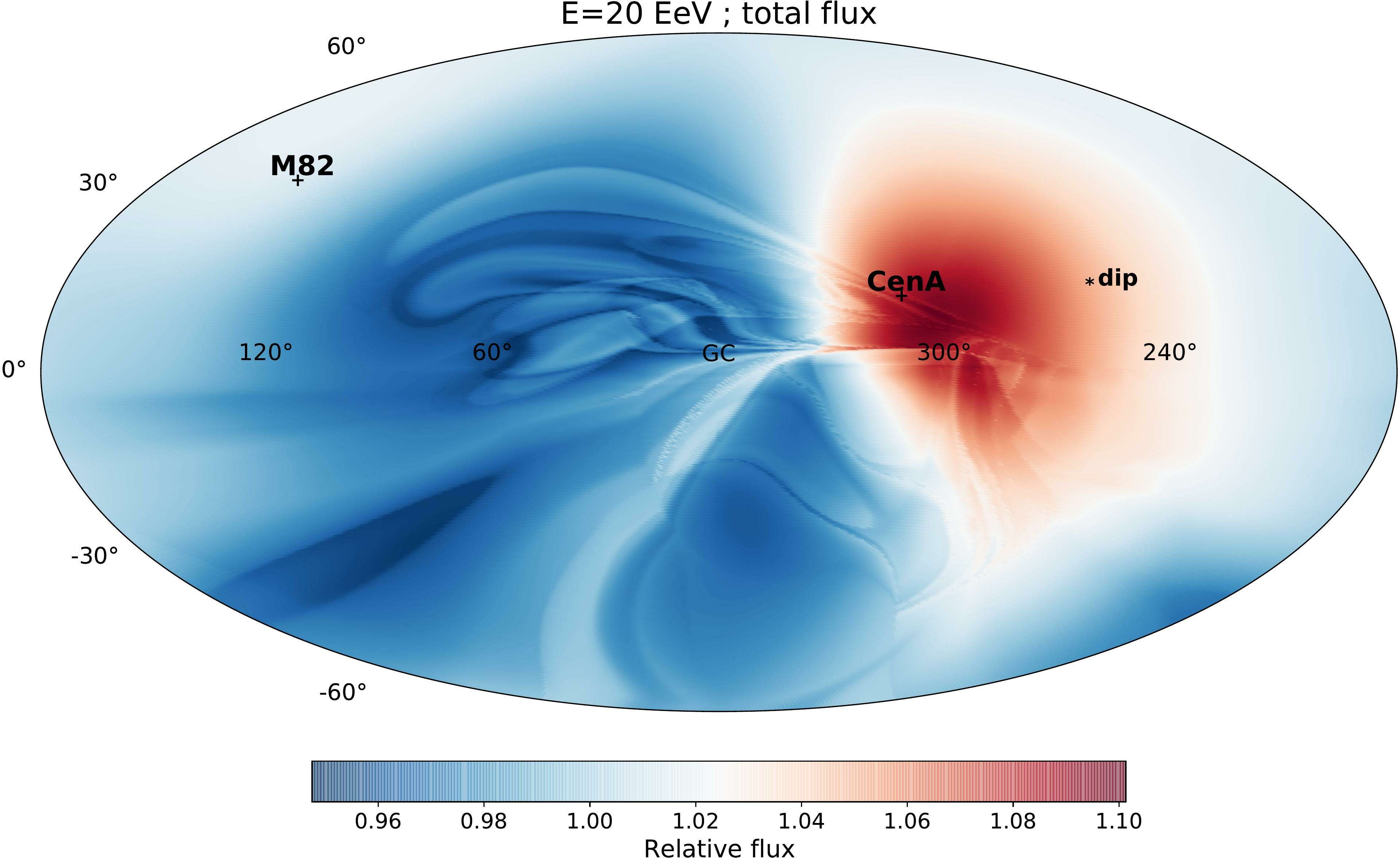}    
    
    \includegraphics[width=0.49\textwidth]{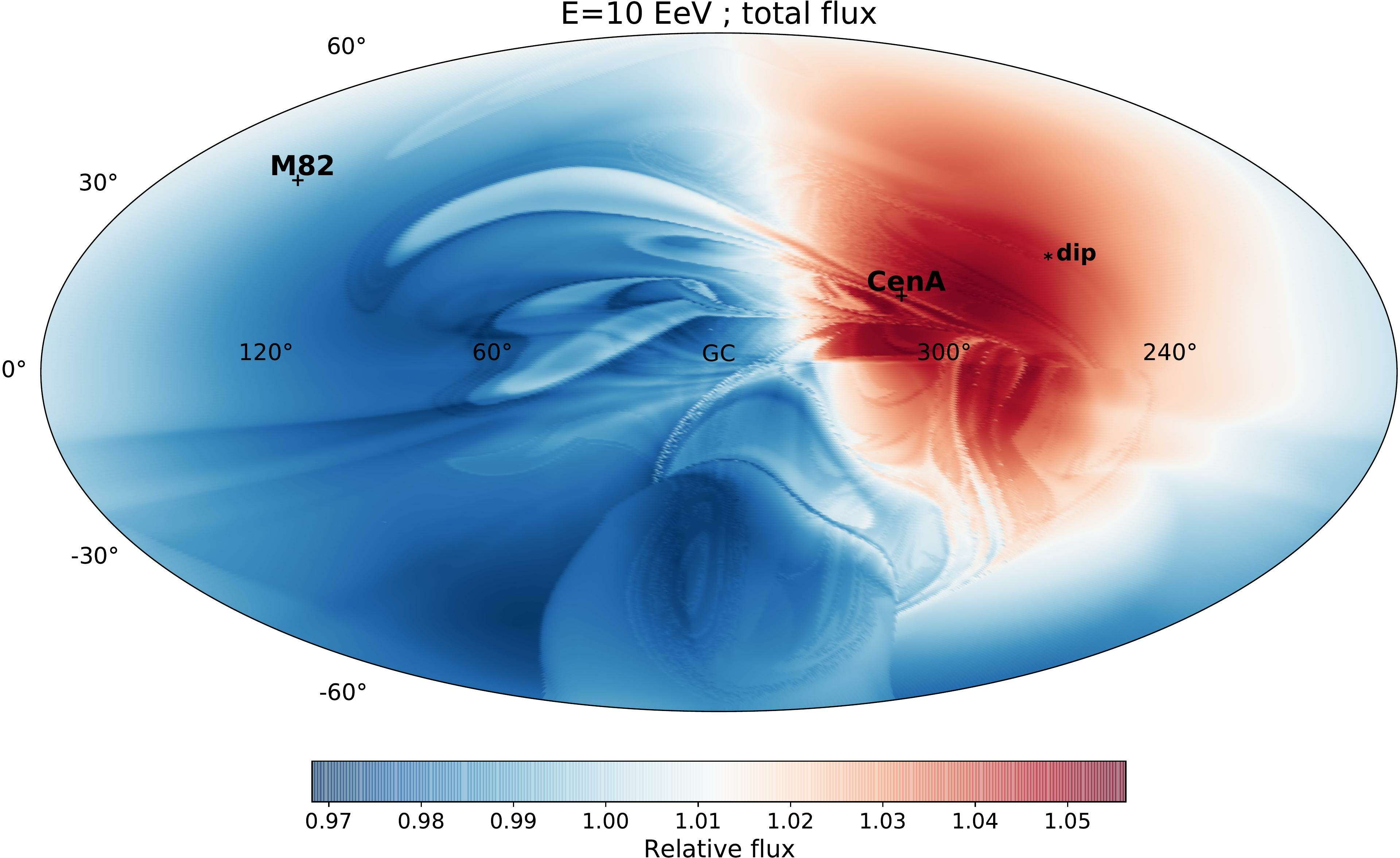} \includegraphics[width=0.49\textwidth]{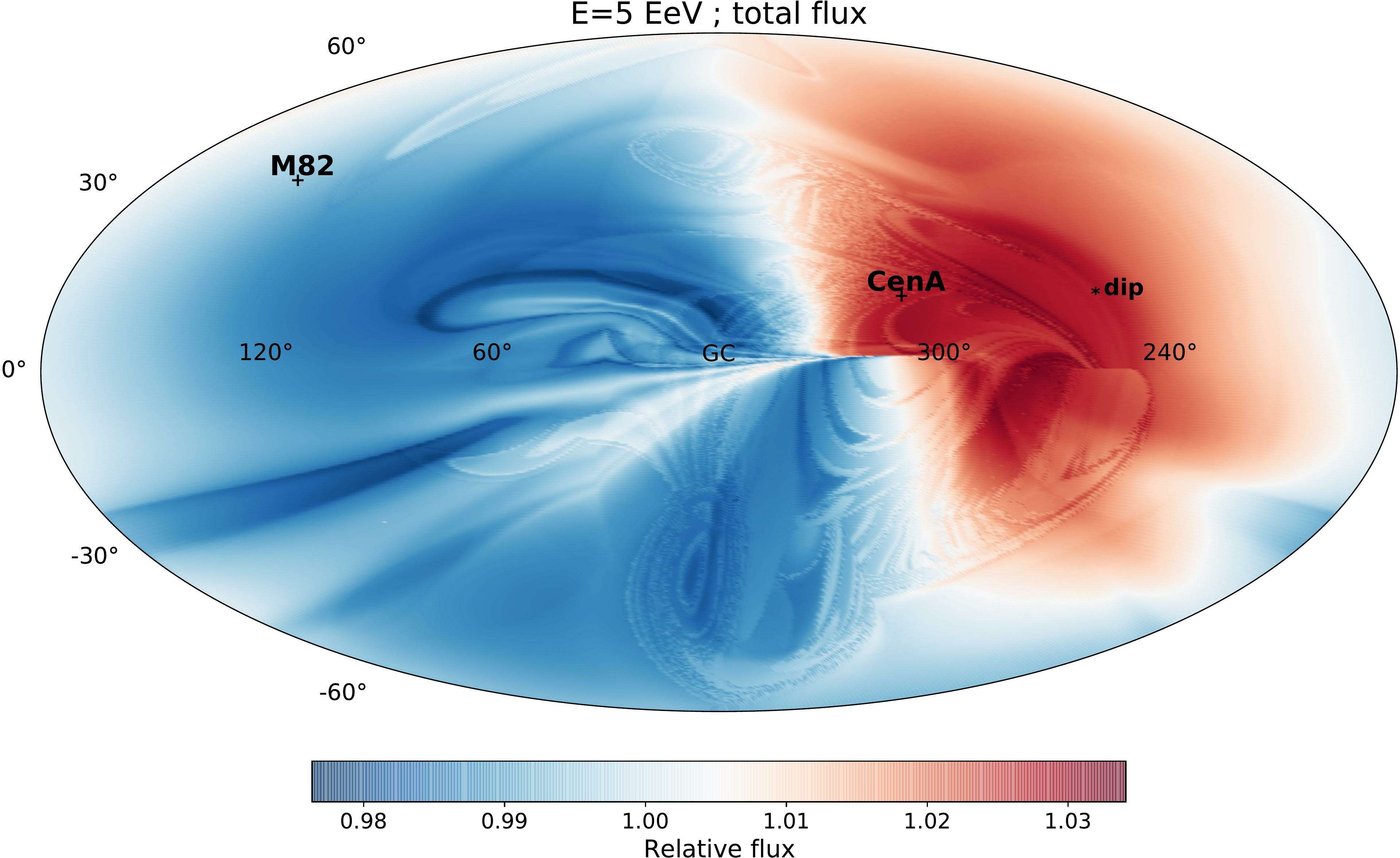}

    \caption{Relative flux maps of CRs emitted by a scenario with mixed composition in which Centaurus~A emits 80\% of the CRs and M81/M82 the other 20\%, with the parameters described in the text. Note the different scales of the relative fluxes in each map.}
    \label{f6}
\end{figure}

The angular distribution of the CR flux outside the halo is obtained through the superposition of the fluxes of the different elements coming from the two sources, as described in Section~2. After accounting for the effects of the Galactic magnetic field, the resulting sky maps are shown in Figure~\ref{f6} for  energies  of 5, 10, 20, 40, 80 and 160~EeV. Here the contribution from the different elements, which have different rigidities at a given observed energy and which have relative proportions which are affected by the spectral modulation induced by the turbulent extragalactic field as well as by the rigidity cutoffs, lead to a total flux inheriting characteristic features from each of the individual elements.  In particular, the more localized excesses around Cen~A get most of their contribution at 10~EeV from H, at 20~EeV from H and He, at 40~EeV from He and N, at 80~EeV from N and at 160~EeV from N and Si (while for the localized excess near M81/M82, given the larger assumed rigidity cutoff the contributions at the different energies come from a slightly lighter composition). 

\begin{figure}[t]
    \centering
    \includegraphics[width=0.49\textwidth]{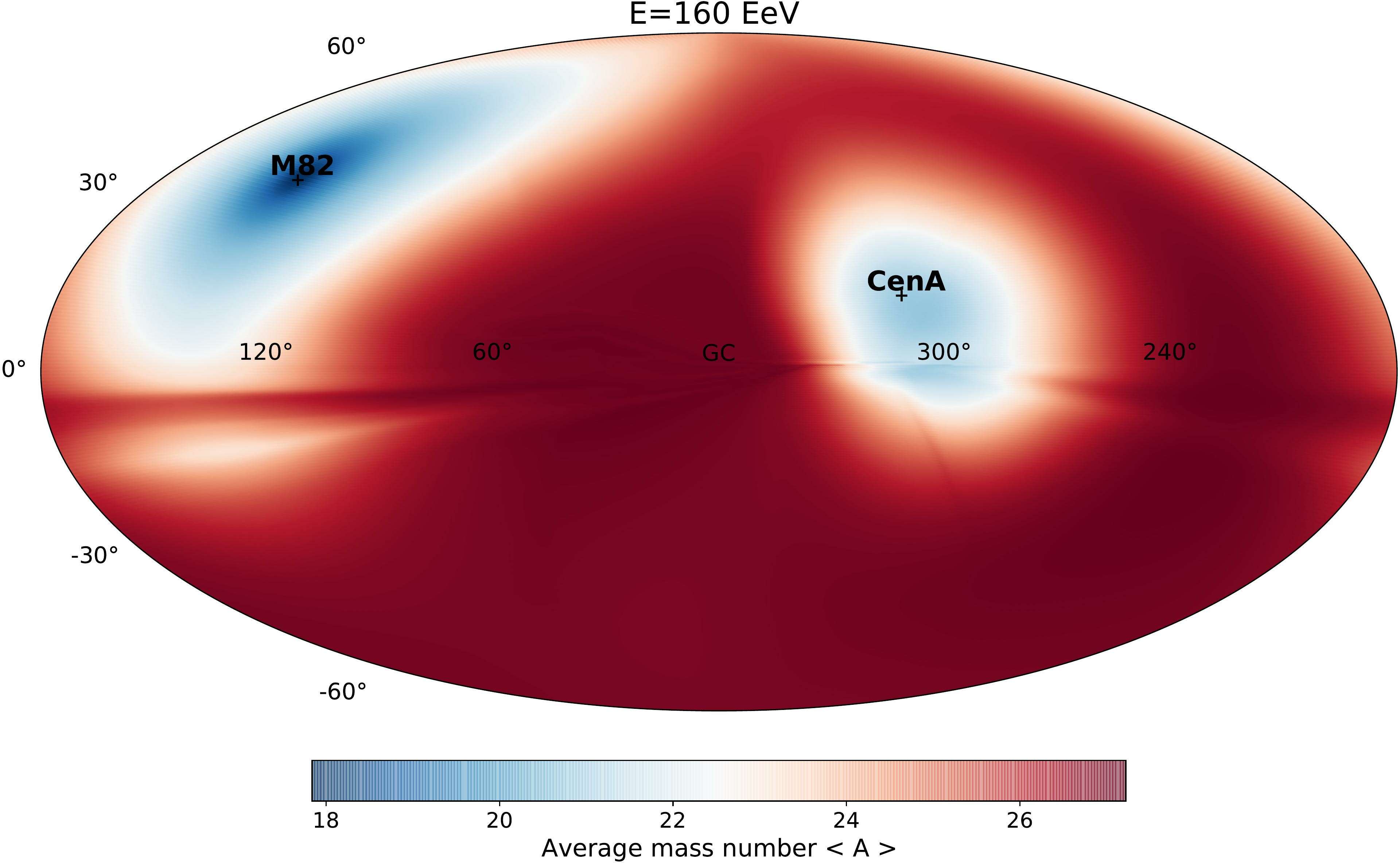} 
    \includegraphics[width=0.49\textwidth]{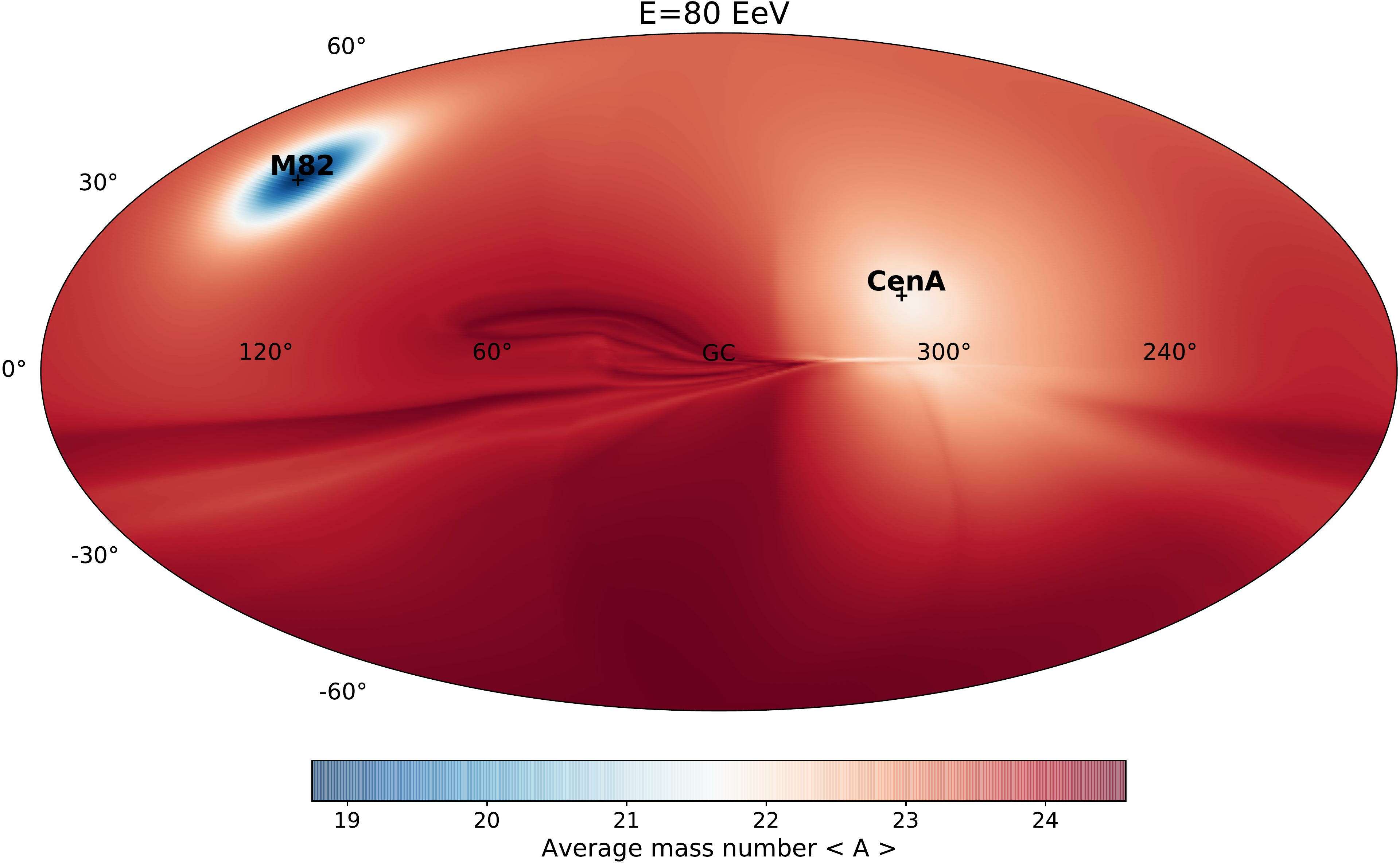} 
    
    \includegraphics[width=0.49\textwidth]{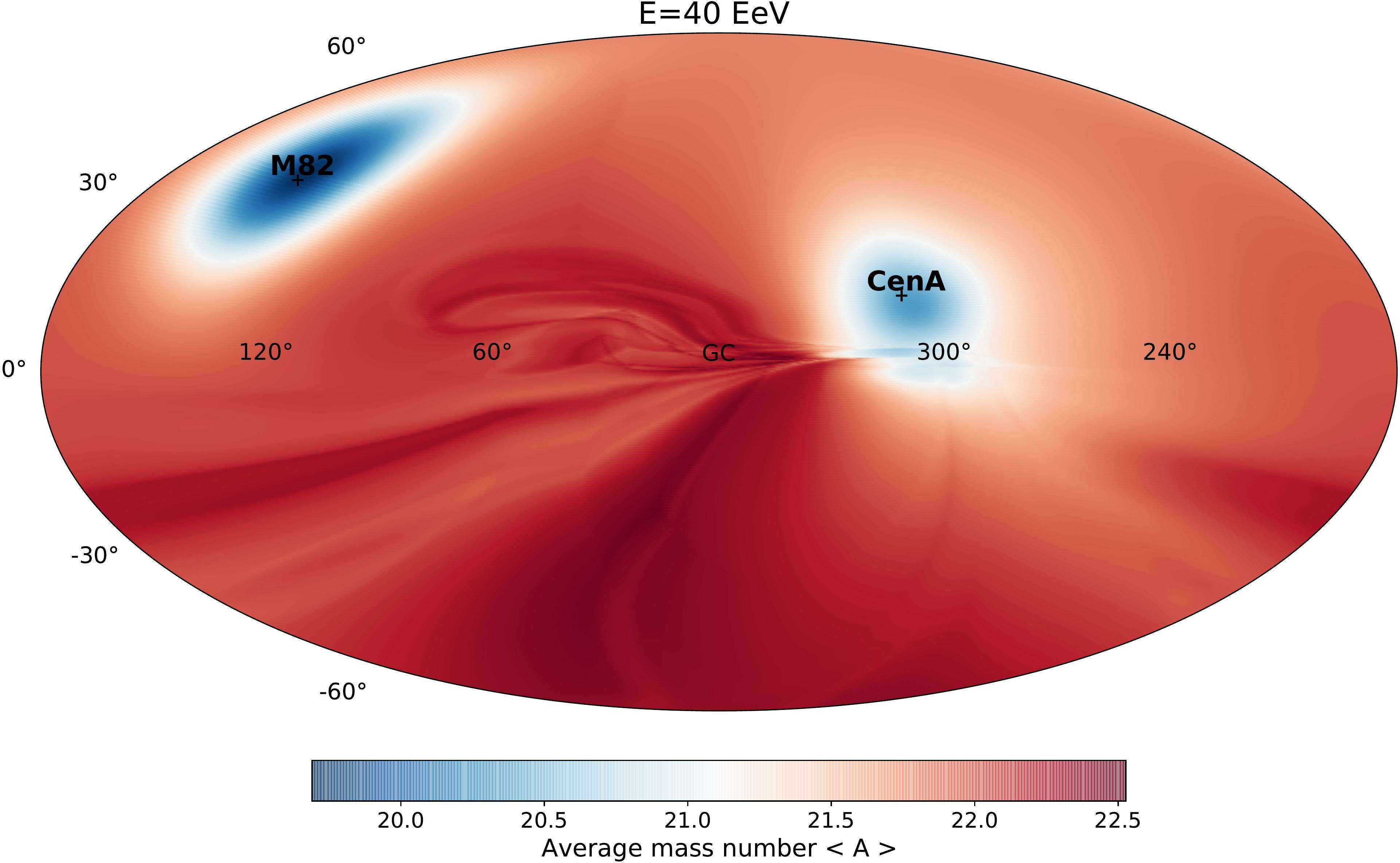} 
    \includegraphics[width=0.49\textwidth]{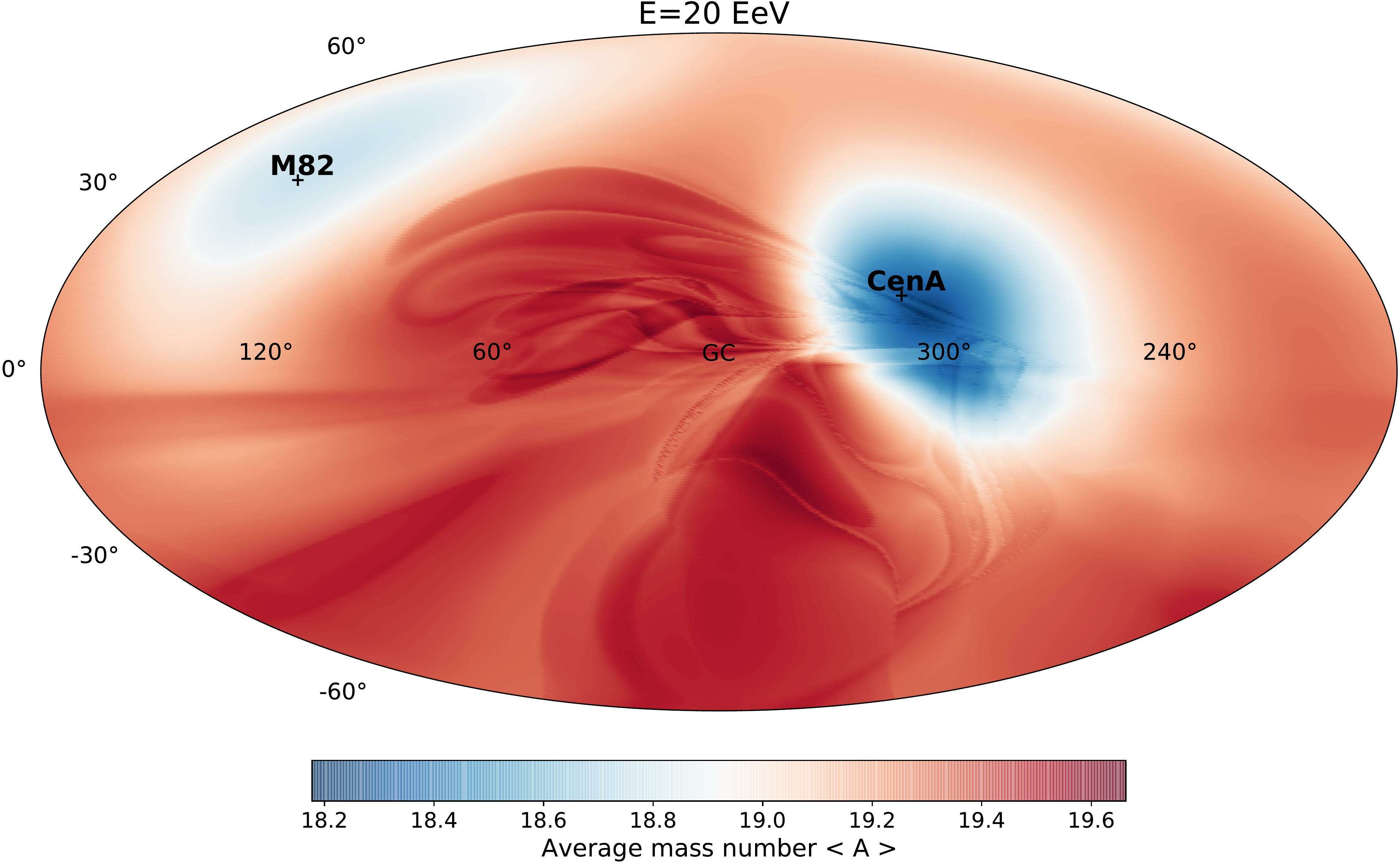}    
    
    \includegraphics[width=0.49\textwidth]{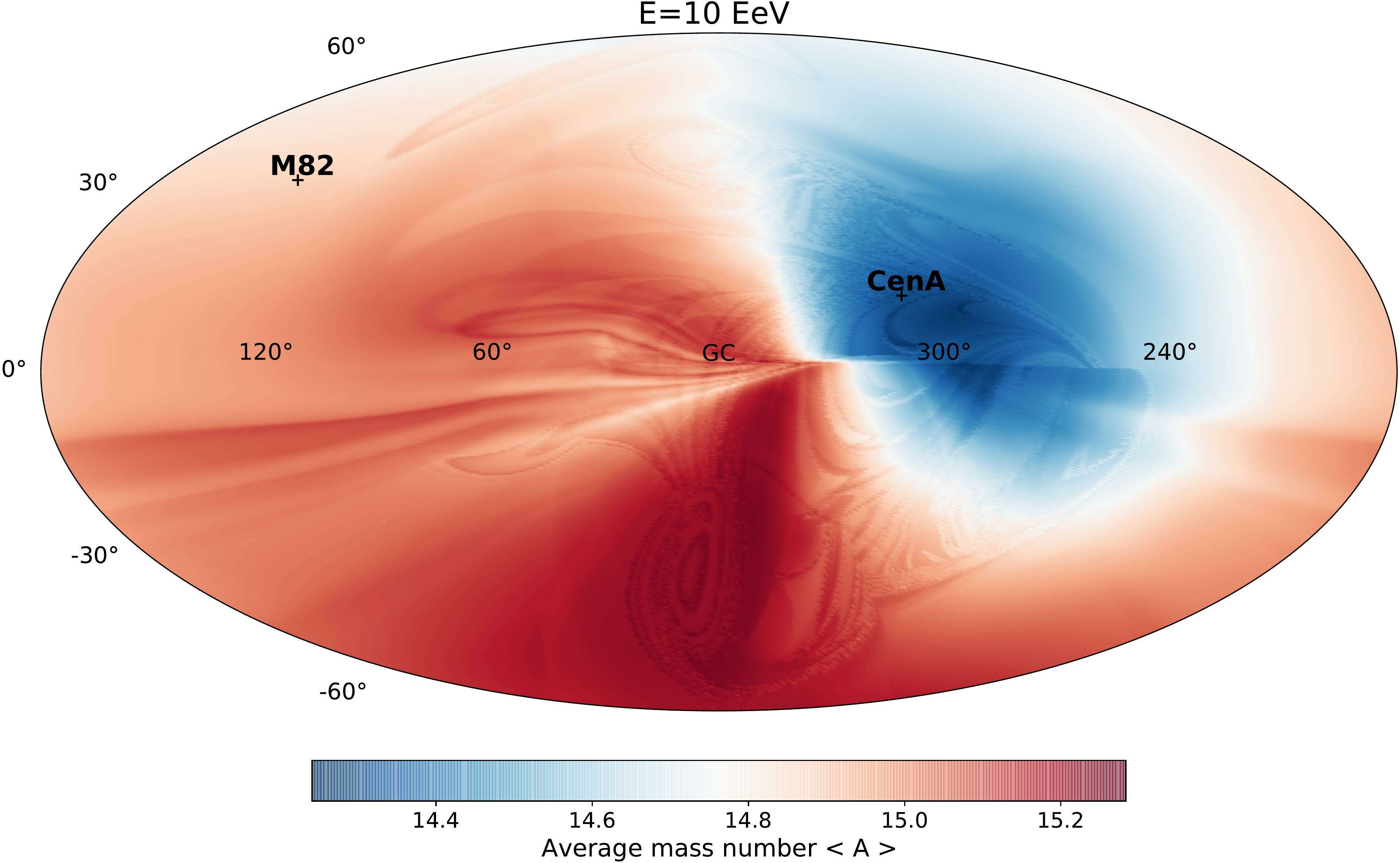} \includegraphics[width=0.49\textwidth]{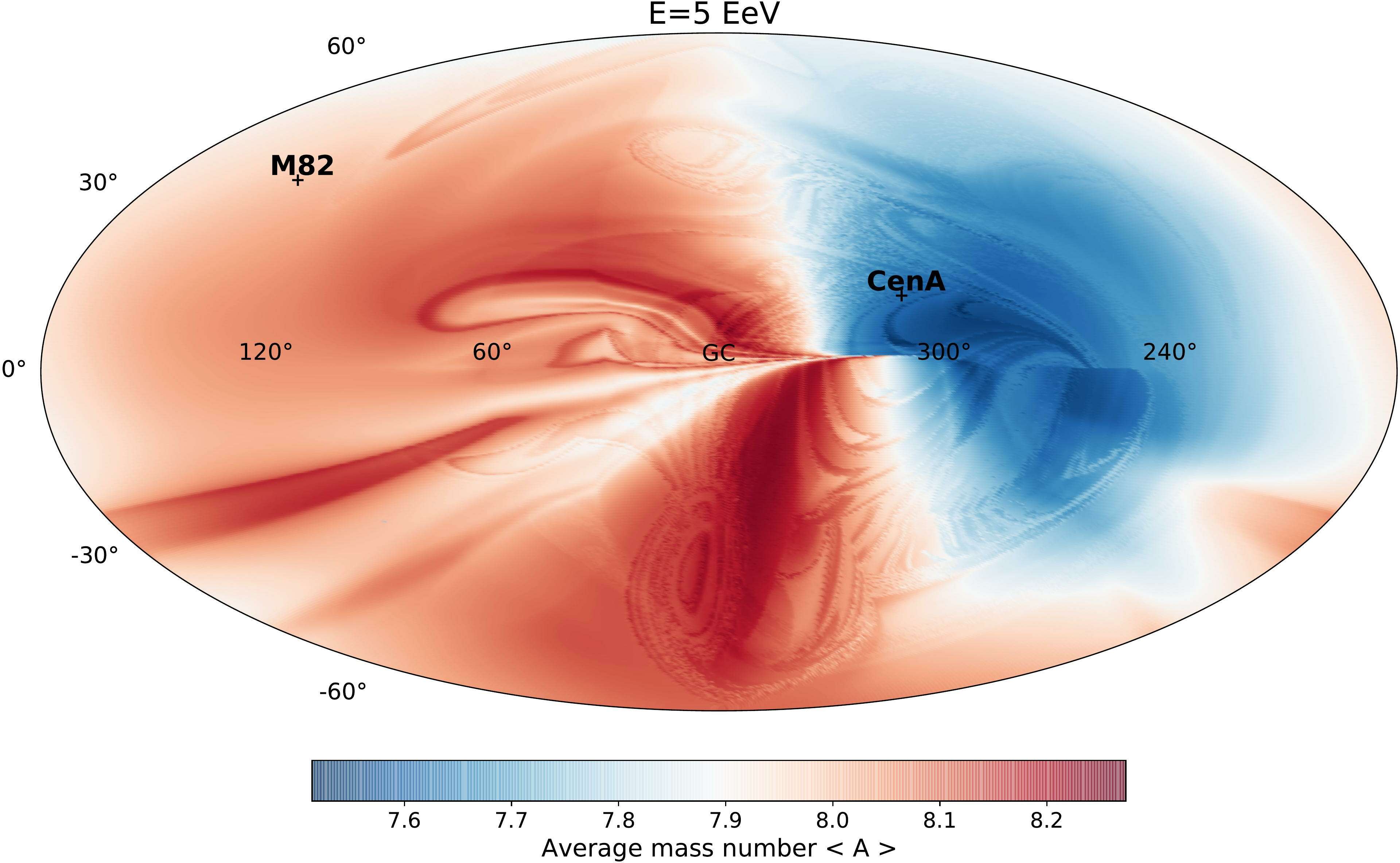}

    \caption{Maps of the average CR mass number for the mixed composition scenario with the two nearby sources described in the text.}
    \label{f7}
\end{figure}

This can be appreciated in Figure~\ref{f7}, where   we show the maps displaying the average value of the mass number $\langle A\rangle$ in the different sky directions for the scenario being considered. The average composition indeed becomes lighter in the regions showing a localized excess in Figure~\ref{f6}.
Also note that due to the changing composition, the scale of the overdensities doesn't necessarily decrease monotonically with increasing energies (e.g., the one around Cen~A at 40~EeV is more concentrated than the one appearing at 80~EeV). One also finds that at the lower energies shown of 5~EeV and 10~EeV, the overall distribution has a strong dipolar modulation, pointing not far from the outer spiral arm of the magnetic field model, as has indeed been observed by the Auger Observatory \cite{science}. However, note that due to the incomplete sky coverage of the existing observatories, the reconstructed dipolar amplitudes in the southern (Auger) and northern (Telescope Array) observatories would differ from the whole sky one. Given that Auger doesn't observe the region around M82 (which lies at declination 69.7$^\circ$), while TA misses a significant part of the flux from Cen~A (at declination -43$^\circ$), and that the two sources are in quite opposite directions in the sky, some reduction in the total dipolar amplitude with respect to those inferred from each separate source (and those measured by each separate detector) could be expected. In particular, for the fluxes shown in Figure~\ref{f6} one infers a dipolar amplitude of 3.4, 4.5 and 7.7\% for $E=10$, 20 and 40~EeV (pointing towards the label `dip' indicated in the maps), while  considering only the Cen~A source these amplitudes would have been 5.2, 7.0 and 11.9\% respectively, and for the M82 case 4.5, 7.9 and 13.5\%. To obtain these results we used that
$\vec\Delta=3\sum \Phi_i \hat n_i/\sum\Phi_i$, with $i$ labelling a uniform distribution of arrival directions $\hat n_i$.

\section{Discussion}

We have not attempted in this work to do a detailed fit to the observations, since this would ultimately depend on the specific Galactic (and extragalactic) magnetic  field models adopted, the detailed history of the emission of the  sources, as well as the composition and spectral properties of each source. However, as was shown above all the main ingredients required to account for the kind of features that have been observed (or hinted) are in principle present in the scenario considered.

In particular, below 20~EeV the main feature in the arrival direction distribution is an almost dipolar distribution with a maximum not far from the direction towards the outer spiral arm, with an amplitude growing with energy from a few percent to about 10 percent. Above 20~EeV the fading light components give rise to more localized overdensities, not only near the source directions but also in other neighbouring regions, as a consequence of the regular magnetic field deflections and, in particular, due to the multiple imaging of the sources. The typical angular scale involved in those localized excesses is that of the turbulent extragalactic magnetic field spreading in eq.~(\ref{eqtheta}). The displacement of the source images with decreasing energies could also affect the spectrum observed around the direction of the sources, since at least the light component giving rise to the localized excess at high energies may be displaced towards a different direction at lower energies. Also note that due to the source rigidity cutoffs the overdensities do not necessarily increase monotonically with energy, since the flux from the lighter component fades away and the heavier ones surface out. In particular,  the relatively large gap between the charge of the CNO group elements and the lighter H and He ones may have the implication that the localized anisotropies present at 40 to 60 EeV may not be necessarily stronger around 100~EeV

Let us finally note that a study of the anisotropies discriminating the fluxes according to the CR composition, as could be achievable for instance with the ongoing upgrade of the Pierre Auger Observatory,  should allow to explore in more detail the predictions obtained here. In this way, one may be able to further test whether the ultrahigh-energy cosmic rays arise from just a few nearby sources or if they are instead due to the cumulative contribution from many of them.

\section*{Acknowledgments}
This work was supported by CONICET (PIP 2015-0369). We are very grateful to Diego Harari for fruitful discussions.

\end{document}